\documentclass[aps,pra,showpacs,amsmath,amssymb,amsfonts,lengthcheck,twocolumn,longbibliography,superscriptaddress]{revtex4-2}
\usepackage{graphicx}
\usepackage{subfigure}
\usepackage{amsthm}
\usepackage{verbatim}
\usepackage{dcolumn}
\usepackage{bm}
\usepackage{epsf}
\usepackage{color}
\usepackage[colorlinks=true,citecolor=blue,linkcolor=blue,urlcolor=blue]{hyperref}%
\usepackage{xcolor}
\usepackage{dsfont}

\newcommand{\bra}[1]{\left\langle #1\right|}
\newcommand{\ket}[1]{\left|#1\right\rangle}

\newcommand{\tr}[1]{\mathrm{tr}\left\{#1\right\}}

\newcommand{\e}[1]{\exp{\left(#1\right)}}

\newcommand{\bla}{bla\\bla\\bla\\bla\\bla}

\newcommand{\mc}[1]{\mathcal{#1}}

\makeatletter
\newcommand{\currentfontsize}{\f@size pt}
\makeatother

\begin{document}

\title{Quantum information scrambling in two-dimensional Bose-Hubbard lattices}
        \author{Devjyoti Tripathy}
\email{dtripathy@umbc.edu}
    \affiliation{Department of Physics, University of Maryland, Baltimore County, Baltimore, MD 21250, USA}
 
	\author{Akram Touil}
 \email{atouil@lanl.gov}
  \affiliation{Theoretical Division, Los Alamos National Laboratory, Los Alamos, New Mexico 87545
}
 \affiliation{Center for Nonlinear Studies, Los Alamos National Laboratory, Los Alamos, New Mexico 87545
}
 
	\author{Bartłomiej Gardas}
\email{bartek.gardas@gmail.com}
\affiliation{Institute of Theoretical and Applied Informatics, Polish Academy of Sciences, Ba{\l}tycka 5, 44-100 Gliwice, Poland}
    
    \author{Sebastian Deffner}
    \email{deffner@umbc.edu}
    \affiliation{Department of Physics, University of Maryland, Baltimore County, Baltimore, MD 21250, USA}
    \affiliation{National Quantum Laboratory, College Park, MD 20740, USA}
    
\date{\today}  
	
\begin{abstract} 
It is a well-understood fact that the transport of excitations throughout a lattice is intimately governed by the underlying structures. Hence, it is only natural to recognize that also the dispersion of information has to depend on the lattice geometry. In the present work, we demonstrate that two-dimensional lattices described by the Bose-Hubbard model exhibit information scrambling for systems as little as two hexagons. However, we also find that the OTOC shows the exponential decay characteristic for quantum chaos only for a judicious choice of local observables. More generally, the OTOC is better described by Gaussian-exponential convolutions, which alludes to the close similarity of information scrambling and decoherence theory.
\end{abstract}

\maketitle

\section{Introduction}

In India they call it \emph{Akuri} or \emph{Bhurji}, and in the Spanish speaking world it is \emph{Revuelto} -- yet all of these terms refer rather literally to one of the most common breakfast items, namely \emph{scrambled eggs} \cite{scrambled_eggs}. Nowadays it has become a typical marketing strategy to embellish pretty much any product name  with the attribute ''quantum", and even ``quantum eggs'' have become a  phenomenon of popular culture \cite{quantum_egg_1,quantum_egg_2}. 

Yet, \emph{quantum information scrambling} \cite{Touil2024EPL} is an actual scientific term that refers to the spread of initially localized quantum information throughout non-local degrees of freedom in complex many body systems \cite{ Touil2021PRX,Roberts2016PRL,Anand2022brotocsquantum,Dong2022PRR,PatrickHayden_2007,Blok2021PRX}. Arguably, the most commonly used quantifier for the rate with which information becomes non-local is the Out-of-Time-Ordered Correlator (OTOC) \cite{Huang2017AP,PhysRevA.95.012120, Swingle2018NP,Hashimoto2017,PhysRevA.107.032818,Roberts2015,PhysRevLett.115.131603,FAN2017707}. In particular, an exponential scaling of the OTOC as a function of time signifies quantumly chaotic dynamics \cite{Swingle2016PRA,COTLER2018318,Hashimoto2017}, with corresponding quantum Lyapunov exponents \cite{Maldacena2016, PhysRevLett.124.140602,PhysRevD.94.106002}.

In recent years, quantum information scrambling has attracted significant interest, see for instance a recent perspective \cite{Touil2024EPL} and references therein. However, a comprehensive analysis of the dynamics of complex many body systems typically requires sophisticated numerical tools. Hence, a large fraction of the literature has focused on systems with effectively one-dimensional geometry, such as the disordered XXX chain  \cite{Iyoda2018PRA} or the mixed-field Ising model \cite{Sahu2019PRL,Sahu2020PRB}. Note that paradigmatic examples of \emph{fast scramblers} are built from Sachdev-Ye-Kitaev (SYK) models \cite{Sachdev1993PRL,Maldacena2016PRD,Rosenhaus2019JPA,Plugge2020PRL}, which, however, do not have a clear notion of dimensionality due to their full connectivity.

In the present work, we study the dynamics of information scrambling in lattices with two-dimensional geometry. As a Hamiltonian, we choose the Bose-Hubbard model, for which chaotic behavior has been reported \cite{Kolovsky2016INMP,Shen2017PRB}. Note that the Bose-Hubbard Model undergoes a second order quantum phase transition from the superfluid phase to the Mott insulator phase \cite{J.K.Freericks_1994,PhysRevB.95.104306,PhysRevB.107.144510,Wang_2016,FAN2017707}. The quantum critical region is strongly interacting, and the energy conserving interactions between the quasi particles in this regime are responsible for the thermalization of the system at strong couplings \cite{Kollath2007PRL,FAN2017707}. Interestingly, it was found in Ref.~\cite{Shen2017PRB} that the dynamics is quantum chaotic at strong couplings and that the Lyapunov exponent displays a maximum around the quantum critical region. Moreover, the eigenvalue statistics, excessive kurtosis of eigenstates, and the eigenstate thermalisation hypothesis have been analyzed~\cite{Goran2023PRE}.

What makes the Bose-Hubbard model particularly interesting is the fact that in hexagonal lattices the two-dimensional system exhibits Dirac points in the energy bands, which describes the motion of effectively relativistic dynamics \cite{Wehling2014AP}. It appears plausible that the dispersion relation governs the rate with which information can be scrambled, and hence it is only natural to study the effect of the lattice geometry on the dynamics of the OTOC in scrambling systems. Indeed, we find that in two-dimensional lattices, the OTOC is sensitive to the neighborhood of support of initial local operators, and that it displays a transition from Gaussian to near exponential decay as we change the configuration of the lattices and/or increase their sizes. Interestingly, when using the OTOC as a scrambling quantifier, it has been argued that decoherence and scrambling dynamics are hardly distinguishable in open systems \cite{Touil2021PRX,PhysRevLett.122.040404}. This is further supported by our current findings, as the OTOC is, indeed, best  described by a convolution of Gaussian and exponential function, which is in full analogy to the decoherence factor~\cite{Yan2022NJP}.

\section{Preliminaries}
We start by establishing notions and notations, and specifying the model.
\subsection{The Bose-Hubbard model} 
The Hubbard model was originally developed to describe strongly-correlated electrons in solids \cite{Arovas2022ARCMP}. As such, creation and annihilation operators were equipped with the fermionic commutation relation. Yet, also the corresponding bosonic version has found widespread applications, in for instance describing optical lattices \cite{Dutta2015RPP}.

The Hamiltonian is usually written as
\begin{equation}
\label{eq:bh}
    H = -J\sum_{\langle i,j\rangle}\left(a_{i}^{\dagger}a_{j} + a_{j}^{\dagger}a_{i}\right) +\frac{U}{2}\sum_{i}n_{i}\left(n_{i}-1\right)   
\end{equation}
where $J$ is the hopping coefficient, $U$ is the on-site potential and $n_{i}=a_{i}^{\dagger}a_{i}$ describes the number of bosons at site $i$. Note that the lattice geometry is entirely encoded in the first sum.

The Bose-Hubbard model exhibits a quantum phase transition.  When the first term in Eq.~\eqref{eq:bh} is dominant, then the bosons can freely hop from one site to another and thus condense into a superfluid phase. On the other hand, when the second term is dominant, then the bosons have to pay a high potential cost to condense into a single site, and thus they prefer to stay at their respective sites resulting in a Mott insulator state. This transition was observed in ultracold atoms in optical lattices \cite{Greiner2002}, and it has been demonstrated that the behavior around the critical point is well-described by the Kibble-Zurek mechanism \cite{Dziarmaga2014,PhysRevB.107.144510,PhysRevA.97.033626,PhysRevA.98.063601}.

In the following analysis, we will study the dynamics of spreading information through different lattice geometries. As a main diagnostics tool, we will be using the OTOC. For a brief discussion of other quantifiers of scrambling, we refer to Appendix~\ref{sec:appA}.

\subsection{Quantifying scrambling -- the OTOC}
In classical Hamiltonian dynamics, chaos can be identified from the exponential growth of the Poisson bracket \cite{Goldstein2002}. Hence, arguably the most prominent tool to diagnose scrambling of quantum information is a closely related quantity -- the OTOC \cite{Swingle2018NP}.

The OTOC is a four-point correlation function that measures the growth of operators in the Heisenberg picture, and it can be written as
\begin{equation}
\label{eq:otoc4pt}
    \text{OTOC}(t)=\bra{\psi}A^{\dagger}(t)B^{\dagger}A(t)B\ket{\psi}
\end{equation} 
where $A,B$ are two local operators, $A(t) = \exp{(iHt)}A\exp{(-iHt)}$ is the time evolved operator in the Heisenberg picture and $H$ is the Hamiltonian describing the system of interest. Moreover, $\ket{\psi}$ is an initial state of the system. Since $A$ and $B$ are initially local operators, usually defined at two different sites on a lattice, they commute with each other. However, with time, $A(t)$ develops support on other sites in the lattice and as a result $[A(t),B]\neq 0$. 

It is easy to see that at $t=0$ we simply have $\text{OTOC}(0)=1$, and that for $t>0$ we observe $\text{OTOC}(t)<1$. For spatial systems, we can associate a velocity with the OTOC, called the butterfly velocity \cite{COTLER2018318,shenker2014black}. This velocity characterises local operators' growth in time. For non-relativistic lattices, the Lieb-Robinson velocity places an upper bound on the size of time-evolved operators and is state independent \cite{Gong2023PRL,Nachtergaele2009,hastings2010locality}. It has been shown that the butterfly velocity is an effective state-dependent Lieb-Robinson velocity \cite{roberts2016lieb}.

\section{Scrambling in 2D}
To analyze the scrambling properties of the Bose-Hubbard model, we solved the ensuing dynamics numerically. To this end, we used Krylov subspace methods \cite{FORD2015491} to compute the time evolution operator. For the present purposes, we chose the following OTOC
\begin{equation}
\label{eq:otoc_bh}
\text{OTOC}(t)=\langle a_{j}^{\dagger}(t)a_{i}^{\dagger}a_{j}(t)a_{i} \rangle_{\psi}
\end{equation}
where $i$ and $j$ are a pair of specific sites in the lattice, and the initial state is chosen to be \emph{all-up}, $\psi = \ket{1,1,1,\dots}$.
In Appendix~\ref{sec:appA} we briefly present results for one-dimensional lattices with periodic boundary conditions. For a lattice with 6 sites and 6 bosons, we observe convincing evidence of scrambling in the superfluid phase.

\begin{figure}  

\centering
  
  \subfigure[ Configuration 1]{\includegraphics[scale=0.1, width= 0.15\textwidth]{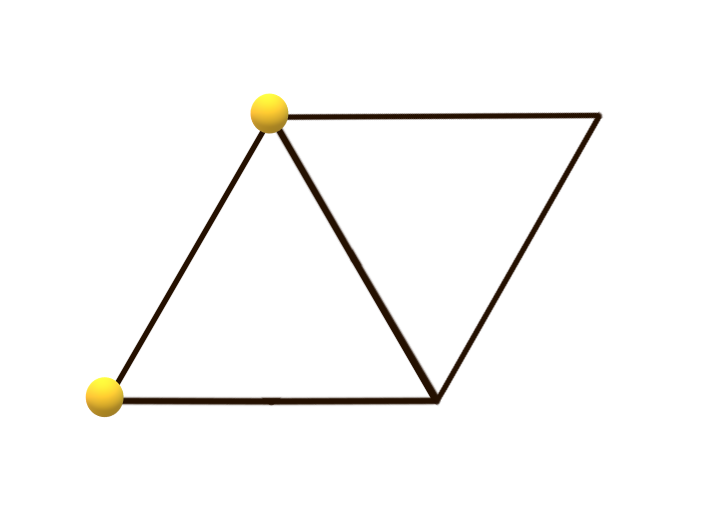}}
  
  \subfigure[ Configuration 2]{\includegraphics[scale=0.1, width=0.15\textwidth]{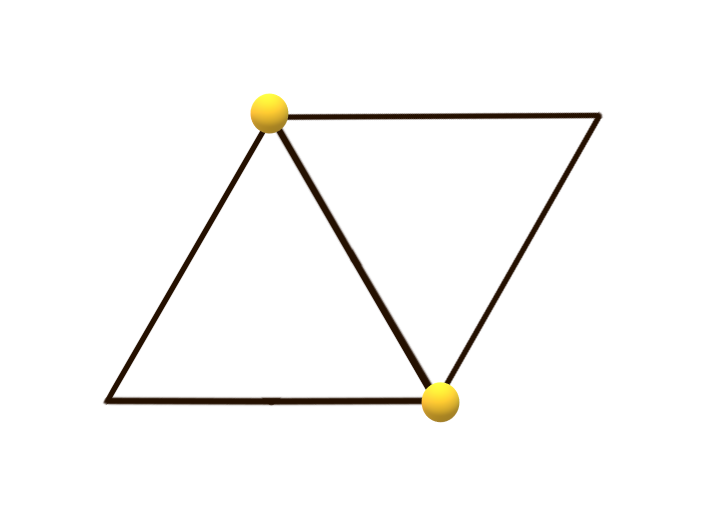}}
\includegraphics[width=.48\textwidth]{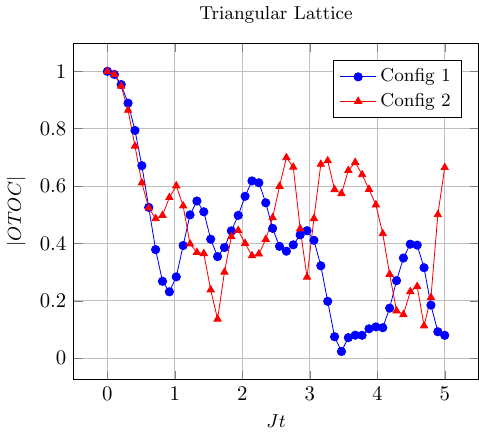}
\caption{OTOC \eqref{eq:otoc_bh} as function of $Jt$, for two configurations of the triangular lattice for $U/J=4$ and $J=4$.}
\label{fig:2dg1}
\end{figure}

\subsection{Triangular and square geometries}
\begin{figure}  

\centering
  
  \subfigure[Configuration 1]{\includegraphics[scale=0.18, width = 0.16\textwidth]{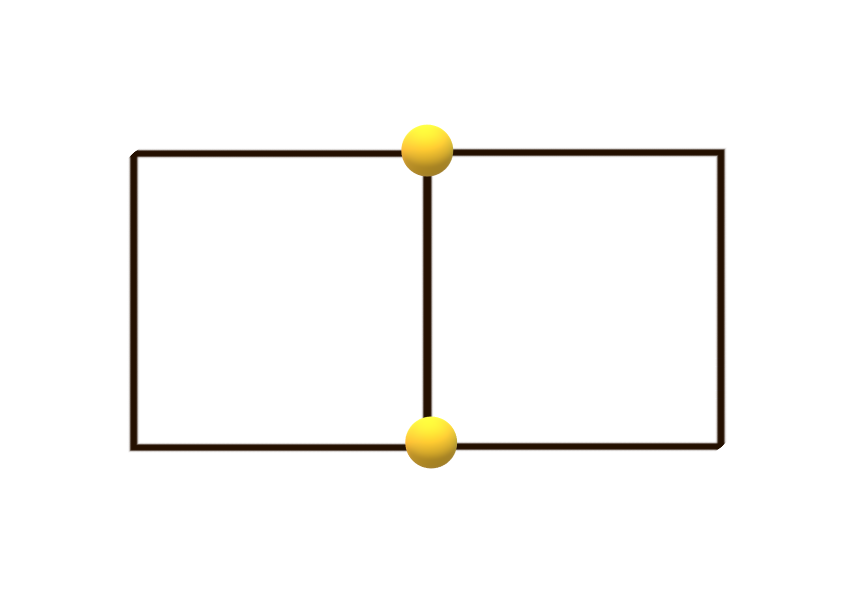}}
  
  \subfigure[Configuration 2]{\includegraphics[scale=0.18, width = 0.16\textwidth]{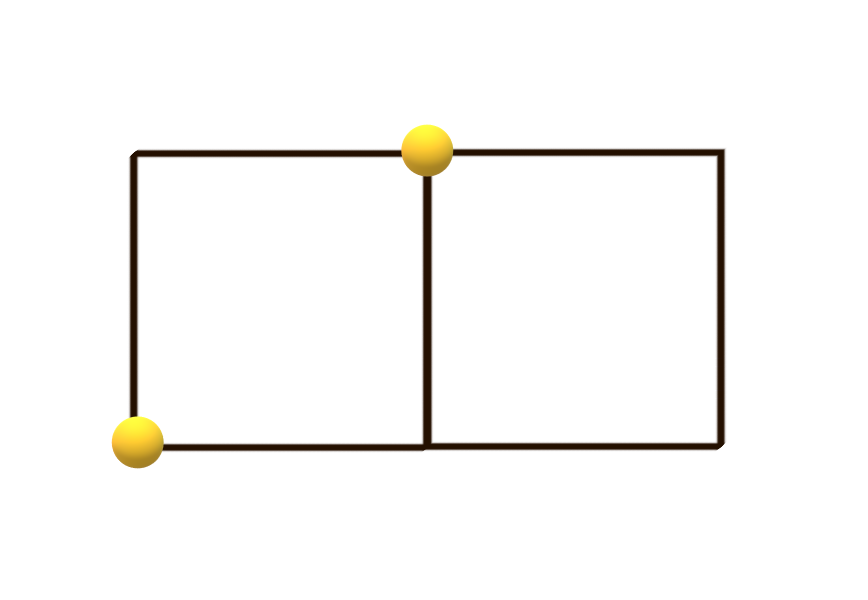}}
  \subfigure[Configuration 3]{\includegraphics[scale=0.18, width = 0.16\textwidth]{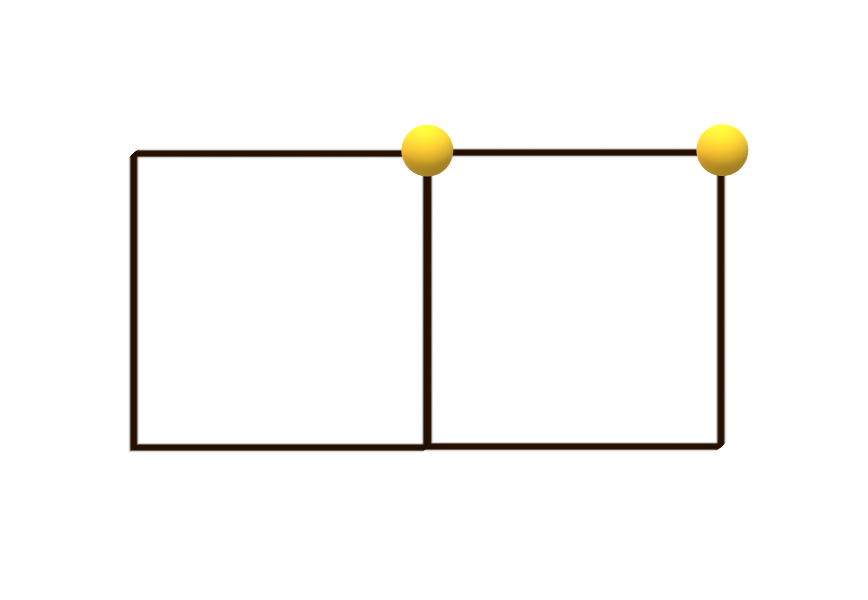}}
\includegraphics[width=.48\textwidth]{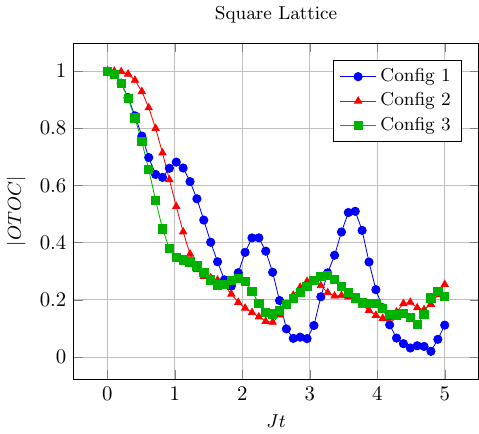}
\caption{OTOC \eqref{eq:otoc_bh} as function of $Jt$, for three configurations of the square lattice for $U/J=4$ and $J=4$..}
\label{fig:2dg4}
\end{figure}

We start with the simplest two-dimensional geometries, before systematically building up to hexagonal geometries.
\paragraph*{Triangular unit cell}
In Fig.~\ref{fig:2dg1} we plot the OTOC \eqref{eq:otoc_bh} for a small lattice comprising two triangles. We observe a rapid decay of the OTOC, followed by sizable fluctuations. This evidences the small system size, which makes an analysis of universal properties of the dynamics inconclusive.
\paragraph*{Square unit cell}
We continue with the next simplest geometry, namely a finite lattice comprised of two squares. The corresponding results are depicted in Fig.~\ref{fig:2dg4}. We observe qualitatively the same behavior as for the triangular cases. Note, however, that the fluctuations from the finite-size effects are significantly less pronounced.

\begin{figure}  

\centering
  
  \subfigure[Configuration 1]{\includegraphics[scale=0.1, width = 0.15\textwidth]{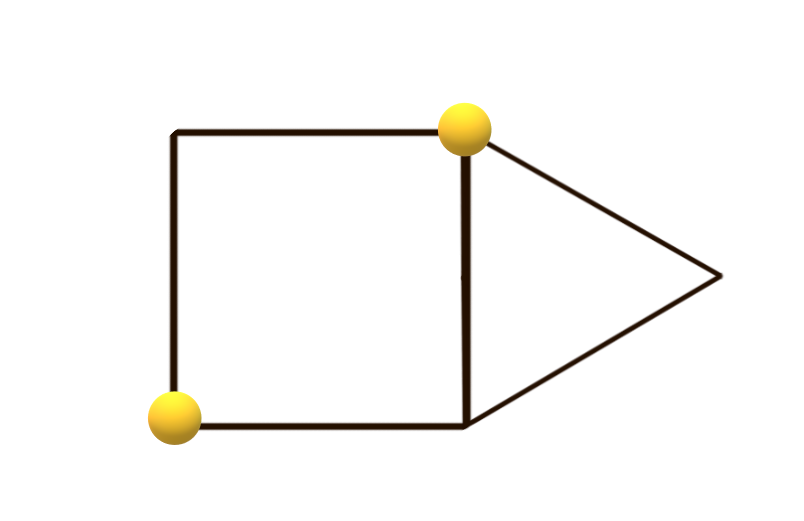}}
  
  \subfigure[Configuration 2]{\includegraphics[scale=0.1, width = 0.15\textwidth]{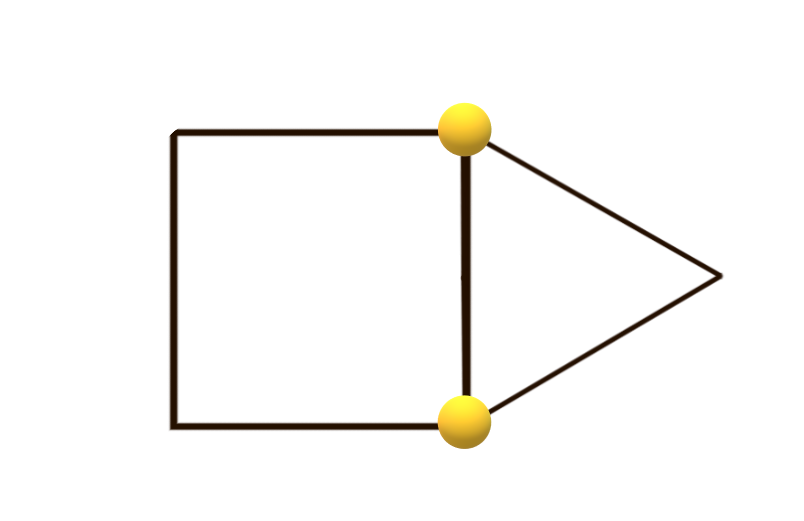}}
  \subfigure[Configuration 3]{\includegraphics[scale=0.1, width = 0.15\textwidth]{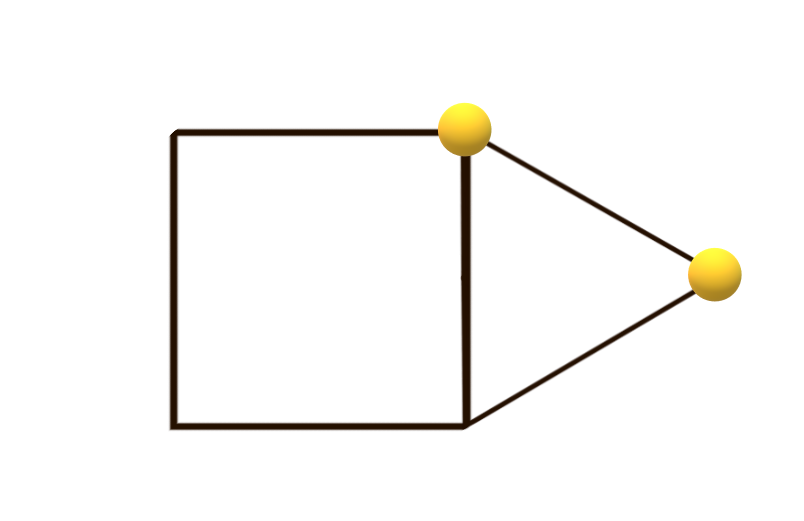}}
\includegraphics[width=.48\textwidth]{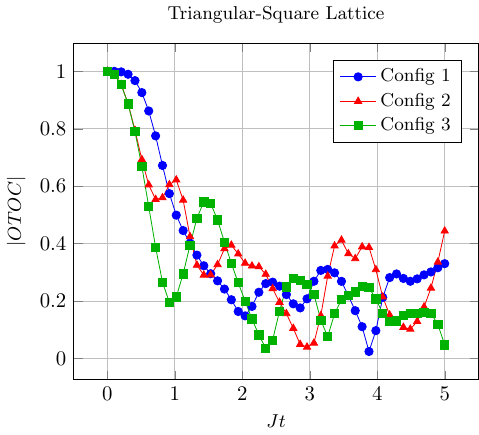 }
\caption{OTOC \eqref{eq:otoc_bh} as function of $Jt$, for three configurations of the triangular-square lattice for $U/J=4$ and $J=4$..}
\label{fig:2dg3}
\end{figure}

\paragraph*{Triangular-square unit cell}
As a final example, before moving on to hexagonal geometries, we consider a mixed case. In Fig.~\ref{fig:2dg3} we summarize findings for lattice configurations that are comprised of a triangle and a square. We find that the increased complexity of the geometry does not lead to qualitatively different behavior, but rather that the dynamics is still governed by finite size effects. In fact, the irregular fluctuations of the OTOC are more pronounced than for the regular square configurations in Fig.~\ref{fig:2dg4}. 

\subsection{Hexagonal geometries}
\begin{figure}
\includegraphics[scale=0.3]{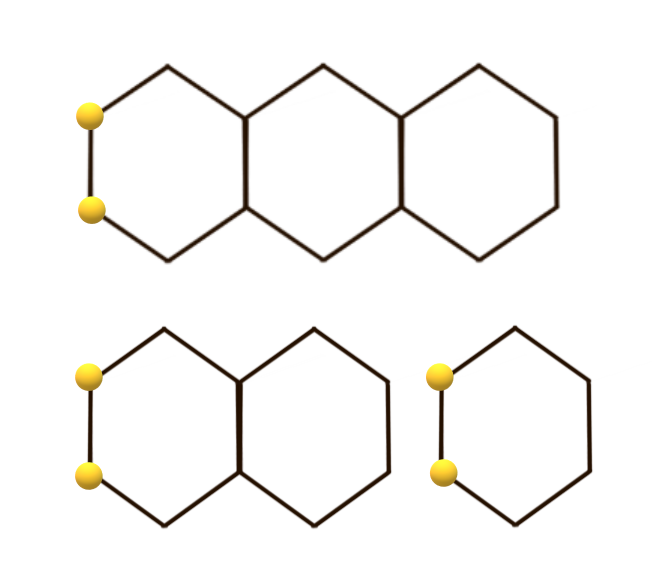}
\caption{Three sizes of the finite hexagonal lattice with neighboring local operators.}
\label{fig:config_a}
\end{figure}

From the simplest geometries discussed so far, we have found that while the OTOC does exhibit the decay characteristic for scrambling, finite size effects govern the dynamics. The situation becomes more interesting for hexagonal geometries. 
\paragraph*{Strip with neighboring local operators}
We start with choosing the local operators to be on neighboring sites, and consider ``strip'' configurations with one, two, and three hexagons. For the ease of notation, we refer to lattices comprised of $n$ hexagons simply as ``$n$ hex". See Fig.~\ref{fig:config_a} for an illustration. The yellow circles in each lattice indicate the support of the initially local operators in the OTOC \eqref{eq:otoc_bh}.

\begin{figure}  
\includegraphics[width=.48\textwidth]{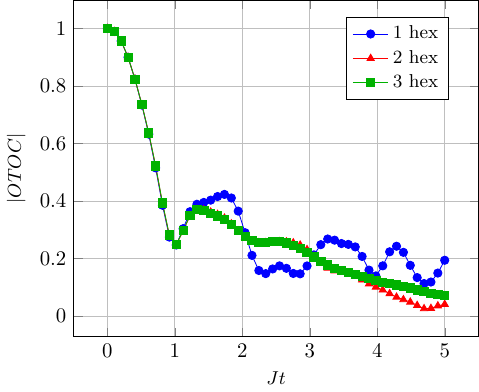}
\caption{OTOC \eqref{eq:otoc_bh} as function of $Jt$ for a strip with neighboring local operators as illustrated in Fig.~\ref{fig:config_a} for $U/J=4$ and $J=4$.}
\label{fig:config_same}
\end{figure}

Our results are summarized in Fig.~\ref{fig:config_same}. Observe that at early times, the decay of the OTOC \eqref{eq:otoc_bh} is independent of the size of the systems. Early in the evolution quantum information remains localized in the first hexagon, and only as time progresses excitations travel father into the lattice. This observation is further supported by the fact that the OTOC for 2 hex departs from the behavior of 3 hex later than the OTOC for 1 hex.
\paragraph*{Strip with distant local operators}
\begin{figure}
\includegraphics[scale=0.3]{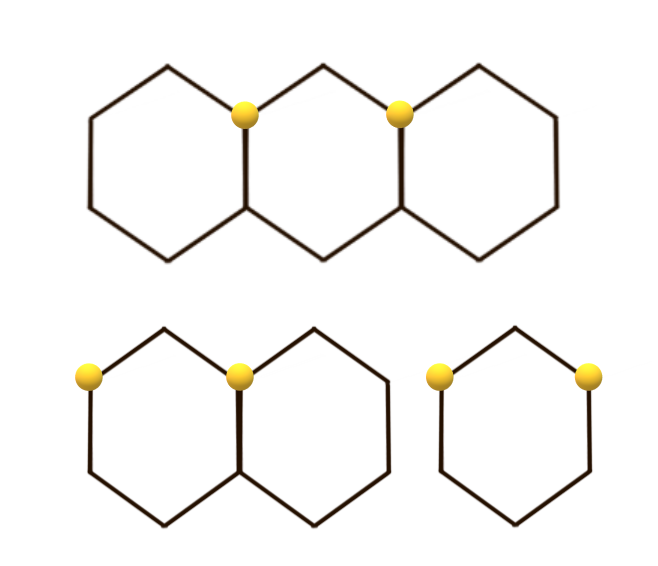}
\caption{Three sizes of the finite hexagonal lattice with distant local operators.}
\label{fig:config_b}
\end{figure}
The observed behavior for a second configuration is similar, while also markedly different. If the initial operators are chosen on ``distant'' lattice sites, as illustrated in Fig.~\ref{fig:config_b}, we again find the early time dynamics to be independent of the size of the system. This is depicted in Fig.~\ref{fig:config_strip}. However, we also notice an earlier departure of the three curves, as well as a much weaker decay of the OTOC at very early times.

\begin{figure}  
\includegraphics[width=.48\textwidth]{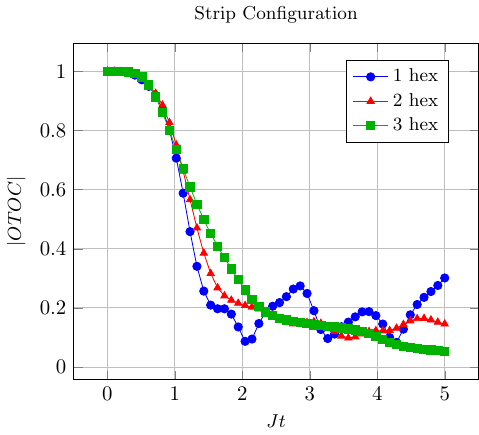}
\caption{OTOC \eqref{eq:otoc_bh} as function of $Jt$ for a strip with distant local operators as illustrated in Fig.~\ref{fig:config_b} for $U/J=4$ and $J=4$.}
\label{fig:config_strip}
\end{figure}

\paragraph*{Bose-Hubbard flakes}

\begin{figure}
\includegraphics[scale=0.3]{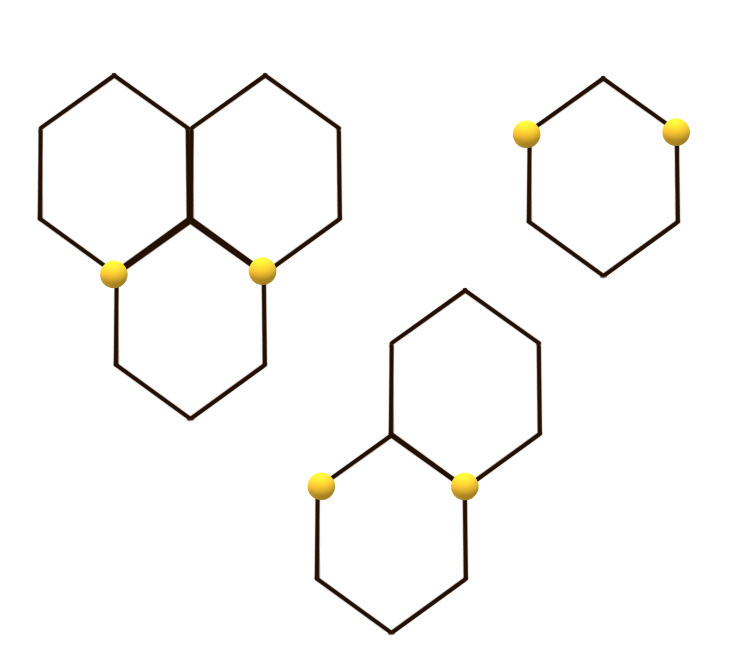}
\caption{Three sizes of the finite hexagonal lattice, including the Bose-Hubbard flake.}
\label{fig:config_c}
\end{figure}

\begin{figure}  
\includegraphics[width=.48\textwidth]{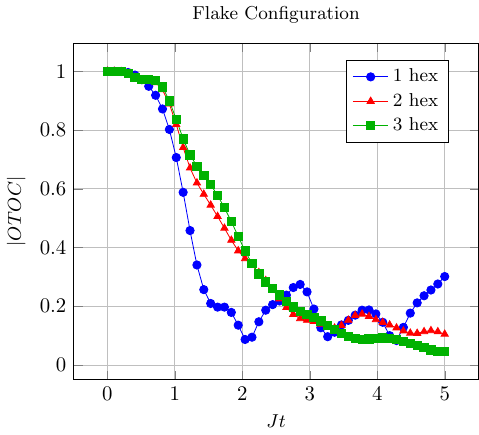}
\caption{OTOC \eqref{eq:otoc_bh} as function of $Jt$ for the Bose-Hubbard flake for distant local operators as illustrated in Fig.~\ref{fig:config_b} for $U/J=4$ and $J=4$.}
\label{fig:config_flake}
\end{figure}

As a third and final example, we solved the dynamics of a ``flake'' configuration with distant local operators, cf. Fig.~\ref{fig:config_c}. The resulting OTOC is depicted in Fig.~\ref{fig:config_flake}. We observe qualitatively similar behavior to the strip configuration with distant local operators, cf. Fig.~\ref{fig:config_strip}.

\subsection{Quantum chaos in the Bose-Hubbard model}

As mentioned above, quantum chaotic dynamics are indicated by an exponential decay of the OTOC \eqref{eq:otoc_bh}, whereas non-chaotic scrambling leads to a slower decay at early times. Thus, we fitted the initial behavior of the OTOC to a simple exponential,
\begin{equation}
\label{eq:exp}
\text{OTOC}(t)\sim \e{\lambda\, \left(t-|x|/v\right)}
\end{equation}
as well as to a Gaussian function,
\begin{equation}
\label{eq:Gauss}
\text{OTOC}(t)\sim \e{\lambda\, \left(t-|x|/v\right)^{2}}\,.
\end{equation}
The rationale for this particular choice will become apparent shortly. Note that $v$ is the butterfly velocity, i.e., the rate with which quasiparticle excitations travel through the lattice.

In Tab.~\ref{tab:fit_strip} we summarize our findings for the strip with distant local operators, cf. Fig.~\ref{fig:config_b}. Interestingly, we find that for $1$ hex and $2$ hex, the Gaussian fit \eqref{eq:Gauss} describes the behavior to much higher accuracy than the exponential fit \eqref{eq:exp}. However, for 3 hex the Gaussian fit predicts \emph{negative} butterfly velocities, which is unphysical. Rather, the exponential fit is a much better approximation.

\begin{table}
\begin{ruledtabular}
\begin{tabular}{c c c c} 
lattice & fit type & $\lambda$ & $v$ \\ [0.5ex] 
 \hline\hline
1 hex & \textbf{Gaussian}  & -26.310 & 14.754 \\ 
 \hline
       & Exponential &   -5.619 &  13.299      \\
 \hline     
 2 hex & \textbf{Gaussian} & -16.352 & 16.389 \\
 \hline
       & Exponential & -16.351 & 8.194      \\
 \hline
 3 hex & Gaussian & -3.072 & \textit{-9.591} \\
 \hline
      & \textbf{Exponential} & -3.734 & 12.551 
\end{tabular}
\end{ruledtabular}
\caption{\label{tab:fit_strip}Fitting parameters for the OTOC \eqref{eq:otoc_bh} resulting from the strip with distant local operators in Fig.~\ref{fig:config_strip}}
\end{table}

Similar results are found for the flake configuration in Fig.~\ref{fig:config_c}. The fitting parameters for the OTOC depicted in Fig.~\ref{fig:config_flake} are summarized in Tab.~\ref{tab:fit_flake}. We find that 2 hex and 3 hex are best described with an exponential fit \eqref{eq:exp}. 

\begin{table}
\begin{ruledtabular}
\begin{tabular}{cccc} 
 lattice & fit type & $\lambda$ & $v$ \\ [0.5ex] 
 \hline\hline
 1 hex & \textbf{Gaussian}  & -26.310 & 14.754 \\ 
 \hline
       & Exponential &   -5.619 &  13.299      \\
 \hline     
 2 hex & Gaussian  & -2.536 & \textit{-9.848} \\ 
 \hline
       & \textbf{Exponential} & -3.024 & 12.218 \\
 \hline
 3 hex & Gaussian & -2.330 & \textit{-9.481} \\
 \hline
    & \textbf{Exponential} & -3.003 & 11.205
\end{tabular}
\end{ruledtabular}
\caption{\label{tab:fit_flake} Fitting parameters for the OTOC \eqref{eq:otoc_bh} resulting from the flake with distant local operators in Fig.~\ref{fig:config_flake}}
\end{table}

Our findings indicate that the dynamics in Bose-Hubbard lattices with hexagonal unit cells becomes chaotic for as little as 2 to 3 hexagons. Moreover, we observe a Gaussian to exponential transition, which strongly reminds of similar observations in decoherence theory \cite{Yan2020PRL}. 

\section{Decoherence vs. Scrambling}

Interestingly, such a Gaussian to exponential transition has been discussed in the literature on the decoherence factor in open quantum systems \cite{Yan2022NJP}. To this end, consider a composite system $\mathcal{SE}$ described by the Hamiltonian 
\begin{equation}
    H_\mc{SE}  = \lambda\sigma_{z}\otimes H_{\mathcal{I}} + H_{\mathcal{E}}\,,
\end{equation} 
where $H_{\mathcal{I,E}}$ acts on the environment and $\sigma_{z}$ is the Pauli Z operator acting on the qubit system $\mathcal{S}$. Initializing $\mc{SE}$  in a product state, $\rho_{\mathcal{SE}}(0)=\rho_{\mathcal{S}}(0)\otimes\rho_{\mathcal{E}}(0)$, the decoherence function $r(t)$ can be written as 
\begin{equation}
\label{eq:deco-le}
 r(t) =\bra{n} \e{iH_{\mathcal{E}}t}\,\e{-i(H_{\mathcal{E}} + \lambda H_{\mathcal{P}})t } \ket{n}
\end{equation} 
where $\ket{n}$ are the eigenstates of $H_{\mathcal{E}}$ and $H_{P}\propto H_{I}$. Note that Eq.~\eqref{eq:deco-le} is a Loschmidt echo \cite{cucchietti2004loschmidt,GORIN200633,PhysRevE.80.046216}.

Yan and Zurek \cite{Yan2022NJP} then showed that Eq.~\eqref{eq:deco-le} can be expressed as a convolution of Gaussian and exponential functions,
\begin{equation}
\label{eq:convul}
\begin{split}
    r(t) &\propto \e{-\tau t/2}* \e{-\sigma^{2}t^{2}/2}\\
    &\propto \sum_{\pm}\e{\frac{\pm\tau t}{2}}\,\text{Erfc}\left(\frac{\tau/2\pm\sigma^{2}t}{\sqrt{2}\sigma}\right)
\end{split}
\end{equation} 
where $*$ denotes convolution and Erfc refers to the error function. In Ref.~\cite{Yan2022NJP}, it is shown explicitly that under rather general assumptions the overlap between the eigenstates of $H_{\mathcal{E}}$ and eigenstates of $H_{\mathcal{E}}+\lambda H_{\mathcal{P}}$ gives rise to a Lorentzian of width $\tau$, whose Fourier transform is the exponential in Eq.~\eqref{eq:convul}.  The Gaussian contribution comes from the spectral density of the Hamiltonian containing local terms, which is assumed to be a Gaussian with standard deviation $\sigma$.

Remarkably, the same authors also showed in Ref.~\cite{Yan2020PRL} that the Haar averaged OTOC \eqref{eq:otoc4pt}  can also be written as a Loschmidt Echo. Hence, it is not far-fetched to realize that also in our present case the behavior of the OTOC \eqref{eq:otoc_bh} should be well-described by the Gausian-exponential convolution \eqref{eq:convul}.

To this end, consider that a small local subsystem, i.e., a small subset of the lattices sites, is designated as system, and the remaining lattices sites as environment. Then, the local operator $A$ is chosen to live on the support of the system, and $B$ has support initially only in the environment. In this picture, it becomes immediately obvious that the OTOC \eqref{eq:otoc_bh} is identical to the decoherence function describing the loss of coherence from the ``system'' into the larger lattice.

Accounting for reflected quasi-particle exciations due to the finite size of the lattice, we write  
\begin{equation}
\label{eq:conv_otoc}
\begin{split}
    \text{OTOC}(t) &= P \e{\frac{-\tau t}{2}}\text{Erfc}\left(\frac{\tau/2 -\sigma^{2}t}{\sqrt{2}\sigma}\right)\\
    &+ Q \e{\frac{\tau t}{2}}\text{Erfc}\left(\frac{\tau/2 +\sigma^{2}t}{\sqrt{2}\sigma}\right)
\end{split}
\end{equation}
where $P$ and $Q$ are free parameters. Using Eq.~\eqref{eq:conv_otoc} we fitted our earlier results again, and the results are summarized in Tabs.~\ref{tab:convul_strip} and \ref{tab:convul_flake}.

\begin{table}
\begin{ruledtabular}
\begin{tabular}{cccc}
 lattice  & $\tau$ & $\sigma$ & $\tau/\sigma$ \\ 
 \hline\hline
1 hex  & 1.822 & 0.122  & 14.934\\ 
 \hline   
2 hex  & 1.707 & 0.163 & 10.472 \\
 \hline
 3 hex  & 0.751 & 1.138  &0.660 
\end{tabular}
\end{ruledtabular}
\caption{\label{tab:convul_strip}
Fitting parameters for the OTOC \eqref{eq:otoc_bh} as  convolution function \eqref{eq:conv_otoc} resulting from the strip with distant local operators in Fig.~\ref{fig:config_strip}.}
\end{table}

\begin{table} 
\begin{ruledtabular}
\begin{tabular}{cccc}
 lattice  & $\tau$ & $\sigma$ & $\tau/\sigma$ \\ 
 \hline\hline
1 hex  & 1.822 & 0.122  & 14.934\\ 
 \hline    
2 hex  & 0.495 & 0.875 & 0.566 \\
 \hline
3 hex  & 0.667 & 0.954 &0.699
\end{tabular}
\end{ruledtabular}
\caption{\label{tab:convul_flake} Fitting parameters for the OTOC \eqref{eq:otoc_bh} as  convolution function \eqref{eq:conv_otoc} resulting from the flake with distant local operators in Fig.~\ref{fig:config_flake}}
\end{table}

Note that when $\tau \gg \sigma$, the convolution function Eq.~\eqref{eq:convul} reduces to a Gaussian whereas when $\tau \ll \sigma$, it reduces to an exponential. We immediately observe that (i) the convolution is a much better description of the OTOC, and (ii) that the more sophisticated fit is consistent with our above results. 

More importantly, we conclude that for two-dimensional Bose-Hubbard lattices, the behavior of the OTOC is remarkably well-described by a decoherence model. This is consistent with earlier findings that highlight the close relationship of information scrambling and decoherence \cite{Touil2021PRX}.

\section{Concluding remarks}

In this paper, we have studied the notion of information scrambling in two-dimensional lattices described by the Bose-Hubbard model. We have found that for as little as 2 hexagons the OTOC shows the characteristic exponential decay indicating quantum chaotic behavior. However, we have also found that the scrambling dynamics is highly sensitive to the choice of local operators and lattice configuration. In particular, for ``flake'' configurations the OTOC decays more akin to a decoherence functions as described by a Gaussian-exponential convolution, rather than exhibiting a chaotic, exponential decay.

\acknowledgements{%
S.D. acknowledges support from the U.S. National Science Foundation under Grant No. DMR-2010127 and the John Templeton Foundation under Grant No. 62422. B.G. acknowledges support from the National Science Center (NCN), Poland, under Project Sonata Bis 10, No. 2020/38/E/ST3/00269. A.T. acknowledges support from the U.S DOE under the LDRD program at Los Alamos.
}

\appendix 

\section{\label{sec:appA} Scrambling in 1D}
For completeness and to verify our numerical approach, we also solved for the dynamics of the one-dimensional Bose-Hubbard model. As for the two-dimensional case, we computed the OTOC \eqref{eq:otoc_bh} and also other quantifiers of scrambling.

\paragraph*{Mutual Information}

In Ref.~\cite{Touil2020QST} it was shown that the mutual information is a thermodynamically well-motivated quantifier of scrambling. For two subsystems $A$ and, $B$ the bipartite mutual information between $A$ and $B$ is given by
\begin{equation}
    I(A:B) = S_{A} + S_{B} - S_{AB}
\end{equation} 
where $S_{X} = -\tr{\rho_{X}\ln{\rho_{X}}}$ is the Von-Neumann entropy of the corresponding subsystem $X$. 

In Ref.~\cite{Touil2020QST}, it was shown that the change in Haar-averaged OTOC is upper bounded by the change in bipartite mutual information:
\begin{equation}
\label{eq:otoc-mi}
    \Delta\langle \text{OTOC}(t) \rangle_{Haar} \leq\Delta I(A:B)(t)
\end{equation} 
where $\Delta\langle \text{OTOC}(t) \rangle_{Haar} = 1 - \langle\text{OTOC}(t)\rangle_{Haar}$ is a monotonically growing function. This means that $\Delta I(A:B)$ also has to be growing in time.

\begin{figure*}
  \centering
  
  \subfigure[]{\includegraphics[width=0.45\linewidth]{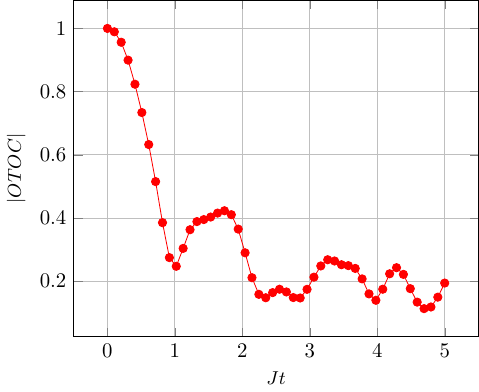 }}
  \hfill
  \subfigure[]{\includegraphics[width=0.45\linewidth]{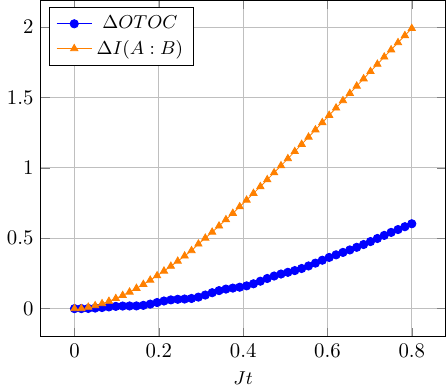}}
  
  \subfigure[]{\includegraphics[width=0.45\linewidth]{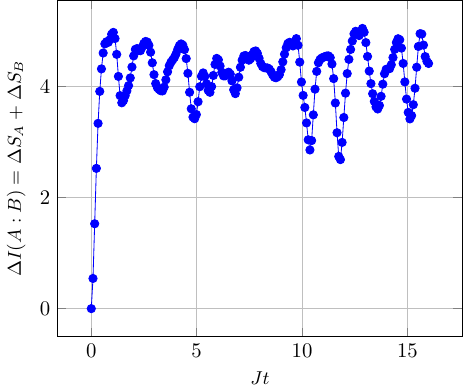}}
  \hfill
  \subfigure[]{\includegraphics[width=0.45\linewidth]{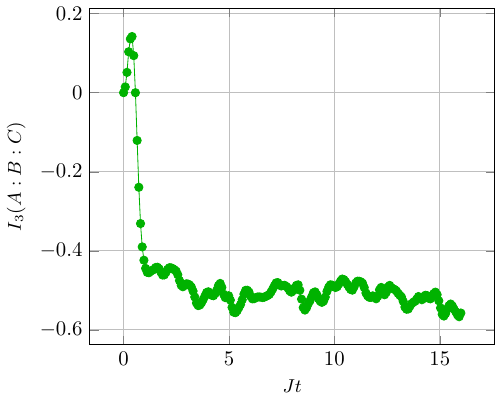}}
  
  \caption{Information Scrambling in the one-dimensional Bose Hubbard model. The scrambling of this model is consistently captured by the bipartite, tripartite mutual information, and the OTOC for $U/J=4$ and $J=4$}
  \label{fig:bh_scrambling}
\end{figure*}

\paragraph*{Tripartite Mutual Information}
Another such information-theoretic quantity that does not depend on the choice of operators is Tripartite Mutual Information (TMI). The Tripartite Mutual Information(TMI) between the three subsystems $A$ and $B$ and $C$ reads
\begin{equation}
    I_{3}(A:B:C) = I(A:B) +  I(A:C) -  I(A:BC)
\end{equation}
Information is said to be scrambled for systems comprised of $B\otimes C$ if TMI becomes negative \cite{Iyoda2018PRA,PhysRevE.98.052205,Hosur2016,e24111532}. This means that the information about $A$ in $BC$ combined has to be greater than the total information about $A$ that $B$ and $C$ have separately.
\paragraph*{Numerical results}
Various measures of scrambling for one hexagonal lattice with 6 sites, 6 bosons and local operators defined over any two nearest neighbor sites on the lattice are shown in Fig.~\ref{fig:bh_scrambling}. We find that the Bose Hubbard model shows information scrambling only when $U\gg J$.

The OTOC shows a rapid decay followed by small oscillations. Mutual Information, over equal bi-partitions of the system, rises rapidly and oscillates about a steady value thereafter. We find that the OTOC for a specific choice of operators (not averaged over) also obeys~\eqref{eq:otoc-mi}. The initial state for TMI is $\ket{\psi}=(\ket{0}_{A}\ket{1}_{B} + \ket{1}_{A}\ket{0}_{B})\ket{1,1}_{C}\ket{1,1,1}_{D} $ where $BCD$ evolves via the Bose-Hubbard Hamiltonian and $A$ remains stationary. The TMI becomes negative and fluctuates about a steady negative value afterward. Therefore, we find that all three quantities indicate the scrambling of quantum information in the Bose Hubbard Model, which confirms the validity of our approach.

\bibliography{ref}

\begin{thebibliography}{62}%
\makeatletter
\providecommand \@ifxundefined [1]{%
 \@ifx{#1\undefined}
}%
\providecommand \@ifnum [1]{%
 \ifnum #1\expandafter \@firstoftwo
 \else \expandafter \@secondoftwo
 \fi
}%
\providecommand \@ifx [1]{%
 \ifx #1\expandafter \@firstoftwo
 \else \expandafter \@secondoftwo
 \fi
}%
\providecommand \natexlab [1]{#1}%
\providecommand \enquote  [1]{``#1''}%
\providecommand \bibnamefont  [1]{#1}%
\providecommand \bibfnamefont [1]{#1}%
\providecommand \citenamefont [1]{#1}%
\providecommand \href@noop [0]{\@secondoftwo}%
\providecommand \href [0]{\begingroup \@sanitize@url \@href}%
\providecommand \@href[1]{\@@startlink{#1}\@@href}%
\providecommand \@@href[1]{\endgroup#1\@@endlink}%
\providecommand \@sanitize@url [0]{\catcode `\\12\catcode `\$12\catcode
  `\&12\catcode `\#12\catcode `\^12\catcode `\_12\catcode `\%12\relax}%
\providecommand \@@startlink[1]{}%
\providecommand \@@endlink[0]{}%
\providecommand \url  [0]{\begingroup\@sanitize@url \@url }%
\providecommand \@url [1]{\endgroup\@href {#1}{\urlprefix }}%
\providecommand \urlprefix  [0]{URL }%
\providecommand \Eprint [0]{\href }%
\providecommand \doibase [0]{https://doi.org/}%
\providecommand \selectlanguage [0]{\@gobble}%
\providecommand \bibinfo  [0]{\@secondoftwo}%
\providecommand \bibfield  [0]{\@secondoftwo}%
\providecommand \translation [1]{[#1]}%
\providecommand \BibitemOpen [0]{}%
\providecommand \bibitemStop [0]{}%
\providecommand \bibitemNoStop [0]{.\EOS\space}%
\providecommand \EOS [0]{\spacefactor3000\relax}%
\providecommand \BibitemShut  [1]{\csname bibitem#1\endcsname}%
\let\auto@bib@innerbib\@empty
\bibitem [{scr(2022)}]{scrambled_eggs}%
  \BibitemOpen
  \href {https://thefoodweknow.com/history-of-scrambled-eggs/} {\bibinfo
  {title} {The history of scrambled eggs}} (\bibinfo {year} {2022})\BibitemShut
  {NoStop}%
\bibitem [{qua(2019)}]{quantum_egg_1}%
  \BibitemOpen
  \href
  {https://blog.qutech.nl/2019/04/21/happy-easter-looking-for-quantum-eggs/}
  {\bibinfo {title} {Happy easter: looking for quantum eggs}} (\bibinfo {year}
  {2019})\BibitemShut {NoStop}%
\bibitem [{qua(2021)}]{quantum_egg_2}%
  \BibitemOpen
  \href {https://egg-inc.fandom.com/wiki/Quantum_Egg} {\bibinfo {title}
  {Quantum egg}} (\bibinfo {year} {2021})\BibitemShut {NoStop}%
\bibitem [{\citenamefont {Touil}\ and\ \citenamefont
  {Deffner}(2024)}]{Touil2024EPL}%
  \BibitemOpen
  \bibfield  {author} {\bibinfo {author} {\bibfnamefont {A.}~\bibnamefont
  {Touil}}\ and\ \bibinfo {author} {\bibfnamefont {S.}~\bibnamefont
  {Deffner}},\ }\bibfield  {title} {\bibinfo {title} {Information scrambling --
  a quantum thermodynamic perspective},\ }\href
  {https://arxiv.org/abs/2401.05305} {\bibfield  {journal} {\bibinfo  {journal}
  {arXiv preprint arXiv:2401.05305}\ } (\bibinfo {year} {2024})}\BibitemShut
  {NoStop}%
\bibitem [{\citenamefont {Touil}\ and\ \citenamefont
  {Deffner}(2021)}]{Touil2021PRX}%
  \BibitemOpen
  \bibfield  {author} {\bibinfo {author} {\bibfnamefont {A.}~\bibnamefont
  {Touil}}\ and\ \bibinfo {author} {\bibfnamefont {S.}~\bibnamefont
  {Deffner}},\ }\bibfield  {title} {\bibinfo {title} {Information scrambling
  versus decoherence---two competing sinks for entropy},\ }\href
  {https://doi.org/10.1103/PRXQuantum.2.010306} {\bibfield  {journal} {\bibinfo
   {journal} {PRX Quantum}\ }\textbf {\bibinfo {volume} {2}},\ \bibinfo {pages}
  {010306} (\bibinfo {year} {2021})}\BibitemShut {NoStop}%
\bibitem [{\citenamefont {Roberts}\ and\ \citenamefont
  {Swingle}(2016{\natexlab{a}})}]{Roberts2016PRL}%
  \BibitemOpen
  \bibfield  {author} {\bibinfo {author} {\bibfnamefont {D.~A.}\ \bibnamefont
  {Roberts}}\ and\ \bibinfo {author} {\bibfnamefont {B.}~\bibnamefont
  {Swingle}},\ }\bibfield  {title} {\bibinfo {title} {Lieb-robinson bound and
  the butterfly effect in quantum field theories},\ }\href
  {https://doi.org/10.1103/PhysRevLett.117.091602} {\bibfield  {journal}
  {\bibinfo  {journal} {Phys. Rev. Lett.}\ }\textbf {\bibinfo {volume} {117}},\
  \bibinfo {pages} {091602} (\bibinfo {year} {2016}{\natexlab{a}})}\BibitemShut
  {NoStop}%
\bibitem [{\citenamefont {Anand}\ and\ \citenamefont
  {Zanardi}(2022)}]{Anand2022brotocsquantum}%
  \BibitemOpen
  \bibfield  {author} {\bibinfo {author} {\bibfnamefont {N.}~\bibnamefont
  {Anand}}\ and\ \bibinfo {author} {\bibfnamefont {P.}~\bibnamefont
  {Zanardi}},\ }\bibfield  {title} {\bibinfo {title} {{BROTOC}s and {Q}uantum
  {I}nformation {S}crambling at {F}inite {T}emperature},\ }\href
  {https://doi.org/10.22331/q-2022-06-27-746} {\bibfield  {journal} {\bibinfo
  {journal} {{Quantum}}\ }\textbf {\bibinfo {volume} {6}},\ \bibinfo {pages}
  {746} (\bibinfo {year} {2022})}\BibitemShut {NoStop}%
\bibitem [{\citenamefont {Yuan}\ \emph {et~al.}(2022)\citenamefont {Yuan},
  \citenamefont {Zhang}, \citenamefont {Wang}, \citenamefont {Duan},\ and\
  \citenamefont {Deng}}]{Dong2022PRR}%
  \BibitemOpen
  \bibfield  {author} {\bibinfo {author} {\bibfnamefont {D.}~\bibnamefont
  {Yuan}}, \bibinfo {author} {\bibfnamefont {S.-Y.}\ \bibnamefont {Zhang}},
  \bibinfo {author} {\bibfnamefont {Y.}~\bibnamefont {Wang}}, \bibinfo {author}
  {\bibfnamefont {L.-M.}\ \bibnamefont {Duan}},\ and\ \bibinfo {author}
  {\bibfnamefont {D.-L.}\ \bibnamefont {Deng}},\ }\bibfield  {title} {\bibinfo
  {title} {Quantum information scrambling in quantum many-body scarred
  systems},\ }\href {https://doi.org/10.1103/PhysRevResearch.4.023095}
  {\bibfield  {journal} {\bibinfo  {journal} {Phys. Rev. Res.}\ }\textbf
  {\bibinfo {volume} {4}},\ \bibinfo {pages} {023095} (\bibinfo {year}
  {2022})}\BibitemShut {NoStop}%
\bibitem [{\citenamefont {Hayden}\ and\ \citenamefont
  {Preskill}(2007)}]{PatrickHayden_2007}%
  \BibitemOpen
  \bibfield  {author} {\bibinfo {author} {\bibfnamefont {P.}~\bibnamefont
  {Hayden}}\ and\ \bibinfo {author} {\bibfnamefont {J.}~\bibnamefont
  {Preskill}},\ }\bibfield  {title} {\bibinfo {title} {Black holes as mirrors:
  quantum information in random subsystems},\ }\href
  {https://doi.org/10.1088/1126-6708/2007/09/120} {\bibfield  {journal}
  {\bibinfo  {journal} {JHEP}\ }\textbf {\bibinfo {volume} {2007}}\bibinfo
  {number} { (09)},\ \bibinfo {pages} {120}}\BibitemShut {NoStop}%
\bibitem [{\citenamefont {Blok}\ \emph {et~al.}(2021)\citenamefont {Blok},
  \citenamefont {Ramasesh}, \citenamefont {Schuster}, \citenamefont {O'Brien},
  \citenamefont {Kreikebaum}, \citenamefont {Dahlen}, \citenamefont {Morvan},
  \citenamefont {Yoshida}, \citenamefont {Yao},\ and\ \citenamefont
  {Siddiqi}}]{Blok2021PRX}%
  \BibitemOpen
\bibfield  {number} {  }\bibfield  {author} {\bibinfo {author} {\bibfnamefont
  {M.~S.}\ \bibnamefont {Blok}}, \bibinfo {author} {\bibfnamefont {V.~V.}\
  \bibnamefont {Ramasesh}}, \bibinfo {author} {\bibfnamefont {T.}~\bibnamefont
  {Schuster}}, \bibinfo {author} {\bibfnamefont {K.}~\bibnamefont {O'Brien}},
  \bibinfo {author} {\bibfnamefont {J.~M.}\ \bibnamefont {Kreikebaum}},
  \bibinfo {author} {\bibfnamefont {D.}~\bibnamefont {Dahlen}}, \bibinfo
  {author} {\bibfnamefont {A.}~\bibnamefont {Morvan}}, \bibinfo {author}
  {\bibfnamefont {B.}~\bibnamefont {Yoshida}}, \bibinfo {author} {\bibfnamefont
  {N.~Y.}\ \bibnamefont {Yao}},\ and\ \bibinfo {author} {\bibfnamefont
  {I.}~\bibnamefont {Siddiqi}},\ }\bibfield  {title} {\bibinfo {title} {Quantum
  information scrambling on a superconducting qutrit processor},\ }\href
  {https://doi.org/10.1103/PhysRevX.11.021010} {\bibfield  {journal} {\bibinfo
  {journal} {Phys. Rev. X}\ }\textbf {\bibinfo {volume} {11}},\ \bibinfo
  {pages} {021010} (\bibinfo {year} {2021})}\BibitemShut {NoStop}%
\bibitem [{\citenamefont {Huang}\ \emph {et~al.}(2017)\citenamefont {Huang},
  \citenamefont {Zhang},\ and\ \citenamefont {Chen}}]{Huang2017AP}%
  \BibitemOpen
  \bibfield  {author} {\bibinfo {author} {\bibfnamefont {Y.}~\bibnamefont
  {Huang}}, \bibinfo {author} {\bibfnamefont {Y.-L.}\ \bibnamefont {Zhang}},\
  and\ \bibinfo {author} {\bibfnamefont {X.}~\bibnamefont {Chen}},\ }\bibfield
  {title} {\bibinfo {title} {Out-of-time-ordered correlators in many-body
  localized systems},\ }\href
  {https://doi.org/https://doi.org/10.1002/andp.201600318} {\bibfield
  {journal} {\bibinfo  {journal} {Ann. Phys.}\ }\textbf {\bibinfo {volume}
  {529}},\ \bibinfo {pages} {1600318} (\bibinfo {year} {2017})}\BibitemShut
  {NoStop}%
\bibitem [{\citenamefont {Yunger~Halpern}(2017)}]{PhysRevA.95.012120}%
  \BibitemOpen
  \bibfield  {author} {\bibinfo {author} {\bibfnamefont {N.}~\bibnamefont
  {Yunger~Halpern}},\ }\bibfield  {title} {\bibinfo {title} {Jarzynski-like
  equality for the out-of-time-ordered correlator},\ }\href
  {https://doi.org/10.1103/PhysRevA.95.012120} {\bibfield  {journal} {\bibinfo
  {journal} {Phys. Rev. A}\ }\textbf {\bibinfo {volume} {95}},\ \bibinfo
  {pages} {012120} (\bibinfo {year} {2017})}\BibitemShut {NoStop}%
\bibitem [{\citenamefont {Swingle}(2018)}]{Swingle2018NP}%
  \BibitemOpen
  \bibfield  {author} {\bibinfo {author} {\bibfnamefont {B.}~\bibnamefont
  {Swingle}},\ }\bibfield  {title} {\bibinfo {title} {Unscrambling the physics
  of out-of-time-order correlators},\ }\href
  {https://doi.org/10.1038/s41567-018-0295-5} {\bibfield  {journal} {\bibinfo
  {journal} {Nat. Phys.}\ }\textbf {\bibinfo {volume} {14}},\ \bibinfo {pages}
  {988} (\bibinfo {year} {2018})}\BibitemShut {NoStop}%
\bibitem [{\citenamefont {Hashimoto}\ \emph {et~al.}(2017)\citenamefont
  {Hashimoto}, \citenamefont {Murata},\ and\ \citenamefont
  {Yoshii}}]{Hashimoto2017}%
  \BibitemOpen
  \bibfield  {author} {\bibinfo {author} {\bibfnamefont {K.}~\bibnamefont
  {Hashimoto}}, \bibinfo {author} {\bibfnamefont {K.}~\bibnamefont {Murata}},\
  and\ \bibinfo {author} {\bibfnamefont {R.}~\bibnamefont {Yoshii}},\
  }\bibfield  {title} {\bibinfo {title} {Out-of-time-order correlators in
  quantum mechanics},\ }\href {https://doi.org/10.1007/JHEP10(2017)138}
  {\bibfield  {journal} {\bibinfo  {journal} {JHEP}\ }\textbf {\bibinfo
  {volume} {2017}}\bibinfo  {number} { (10)},\ \bibinfo {pages}
  {138}}\BibitemShut {NoStop}%
\bibitem [{\citenamefont {Li}\ \emph {et~al.}(2023)\citenamefont {Li},
  \citenamefont {Halperin}, \citenamefont {Wang},\ and\ \citenamefont
  {Bohn}}]{PhysRevA.107.032818}%
  \BibitemOpen
\bibfield  {number} {  }\bibfield  {author} {\bibinfo {author} {\bibfnamefont
  {H.}~\bibnamefont {Li}}, \bibinfo {author} {\bibfnamefont {E.}~\bibnamefont
  {Halperin}}, \bibinfo {author} {\bibfnamefont {R.~R.~W.}\ \bibnamefont
  {Wang}},\ and\ \bibinfo {author} {\bibfnamefont {J.~L.}\ \bibnamefont
  {Bohn}},\ }\bibfield  {title} {\bibinfo {title} {Out-of-time-order correlator
  for the van der waals potential},\ }\href
  {https://doi.org/10.1103/PhysRevA.107.032818} {\bibfield  {journal} {\bibinfo
   {journal} {Phys. Rev. A}\ }\textbf {\bibinfo {volume} {107}},\ \bibinfo
  {pages} {032818} (\bibinfo {year} {2023})}\BibitemShut {NoStop}%
\bibitem [{\citenamefont {Roberts}\ \emph {et~al.}(2015)\citenamefont
  {Roberts}, \citenamefont {Stanford},\ and\ \citenamefont
  {Susskind}}]{Roberts2015}%
  \BibitemOpen
  \bibfield  {author} {\bibinfo {author} {\bibfnamefont {D.~A.}\ \bibnamefont
  {Roberts}}, \bibinfo {author} {\bibfnamefont {D.}~\bibnamefont {Stanford}},\
  and\ \bibinfo {author} {\bibfnamefont {L.}~\bibnamefont {Susskind}},\
  }\bibfield  {title} {\bibinfo {title} {Localized shocks},\ }\href
  {https://doi.org/10.1007/JHEP03(2015)051} {\bibfield  {journal} {\bibinfo
  {journal} {JHEP}\ }\textbf {\bibinfo {volume} {2015}}\bibinfo  {number} {
  (3)},\ \bibinfo {pages} {51}}\BibitemShut {NoStop}%
\bibitem [{\citenamefont {Roberts}\ and\ \citenamefont
  {Stanford}(2015)}]{PhysRevLett.115.131603}%
  \BibitemOpen
\bibfield  {number} {  }\bibfield  {author} {\bibinfo {author} {\bibfnamefont
  {D.~A.}\ \bibnamefont {Roberts}}\ and\ \bibinfo {author} {\bibfnamefont
  {D.}~\bibnamefont {Stanford}},\ }\bibfield  {title} {\bibinfo {title}
  {Diagnosing chaos using four-point functions in two-dimensional conformal
  field theory},\ }\href {https://doi.org/10.1103/PhysRevLett.115.131603}
  {\bibfield  {journal} {\bibinfo  {journal} {Phys. Rev. Lett.}\ }\textbf
  {\bibinfo {volume} {115}},\ \bibinfo {pages} {131603} (\bibinfo {year}
  {2015})}\BibitemShut {NoStop}%
\bibitem [{\citenamefont {Fan}\ \emph {et~al.}(2017)\citenamefont {Fan},
  \citenamefont {Zhang}, \citenamefont {Shen},\ and\ \citenamefont
  {Zhai}}]{FAN2017707}%
  \BibitemOpen
  \bibfield  {author} {\bibinfo {author} {\bibfnamefont {R.}~\bibnamefont
  {Fan}}, \bibinfo {author} {\bibfnamefont {P.}~\bibnamefont {Zhang}}, \bibinfo
  {author} {\bibfnamefont {H.}~\bibnamefont {Shen}},\ and\ \bibinfo {author}
  {\bibfnamefont {H.}~\bibnamefont {Zhai}},\ }\bibfield  {title} {\bibinfo
  {title} {Out-of-time-order correlation for many-body localization},\ }\href
  {https://doi.org/https://doi.org/10.1016/j.scib.2017.04.011} {\bibfield
  {journal} {\bibinfo  {journal} {Science Bulletin}\ }\textbf {\bibinfo
  {volume} {62}},\ \bibinfo {pages} {707} (\bibinfo {year} {2017})}\BibitemShut
  {NoStop}%
\bibitem [{\citenamefont {Swingle}\ \emph {et~al.}(2016)\citenamefont
  {Swingle}, \citenamefont {Bentsen}, \citenamefont {Schleier-Smith},\ and\
  \citenamefont {Hayden}}]{Swingle2016PRA}%
  \BibitemOpen
  \bibfield  {author} {\bibinfo {author} {\bibfnamefont {B.}~\bibnamefont
  {Swingle}}, \bibinfo {author} {\bibfnamefont {G.}~\bibnamefont {Bentsen}},
  \bibinfo {author} {\bibfnamefont {M.}~\bibnamefont {Schleier-Smith}},\ and\
  \bibinfo {author} {\bibfnamefont {P.}~\bibnamefont {Hayden}},\ }\bibfield
  {title} {\bibinfo {title} {Measuring the scrambling of quantum information},\
  }\href {https://doi.org/10.1103/PhysRevA.94.040302} {\bibfield  {journal}
  {\bibinfo  {journal} {Phys. Rev. A}\ }\textbf {\bibinfo {volume} {94}},\
  \bibinfo {pages} {040302} (\bibinfo {year} {2016})}\BibitemShut {NoStop}%
\bibitem [{\citenamefont {Cotler}\ \emph {et~al.}(2018)\citenamefont {Cotler},
  \citenamefont {Ding},\ and\ \citenamefont {Penington}}]{COTLER2018318}%
  \BibitemOpen
  \bibfield  {author} {\bibinfo {author} {\bibfnamefont {J.~S.}\ \bibnamefont
  {Cotler}}, \bibinfo {author} {\bibfnamefont {D.}~\bibnamefont {Ding}},\ and\
  \bibinfo {author} {\bibfnamefont {G.~R.}\ \bibnamefont {Penington}},\
  }\bibfield  {title} {\bibinfo {title} {Out-of-time-order operators and the
  butterfly effect},\ }\href
  {https://doi.org/https://doi.org/10.1016/j.aop.2018.07.020} {\bibfield
  {journal} {\bibinfo  {journal} {Ann. Phys.}\ }\textbf {\bibinfo {volume}
  {396}},\ \bibinfo {pages} {318} (\bibinfo {year} {2018})}\BibitemShut
  {NoStop}%
\bibitem [{\citenamefont {Maldacena}\ \emph {et~al.}(2016)\citenamefont
  {Maldacena}, \citenamefont {Shenker},\ and\ \citenamefont
  {Stanford}}]{Maldacena2016}%
  \BibitemOpen
  \bibfield  {author} {\bibinfo {author} {\bibfnamefont {J.}~\bibnamefont
  {Maldacena}}, \bibinfo {author} {\bibfnamefont {S.~H.}\ \bibnamefont
  {Shenker}},\ and\ \bibinfo {author} {\bibfnamefont {D.}~\bibnamefont
  {Stanford}},\ }\bibfield  {title} {\bibinfo {title} {A bound on chaos},\
  }\href {https://doi.org/10.1007/JHEP08(2016)106} {\bibfield  {journal}
  {\bibinfo  {journal} {JHEP}\ }\textbf {\bibinfo {volume} {2016}}\bibinfo
  {number} { (8)},\ \bibinfo {pages} {106}}\BibitemShut {NoStop}%
\bibitem [{\citenamefont {Xu}\ \emph {et~al.}(2020)\citenamefont {Xu},
  \citenamefont {Scaffidi},\ and\ \citenamefont
  {Cao}}]{PhysRevLett.124.140602}%
  \BibitemOpen
\bibfield  {number} {  }\bibfield  {author} {\bibinfo {author} {\bibfnamefont
  {T.}~\bibnamefont {Xu}}, \bibinfo {author} {\bibfnamefont {T.}~\bibnamefont
  {Scaffidi}},\ and\ \bibinfo {author} {\bibfnamefont {X.}~\bibnamefont
  {Cao}},\ }\bibfield  {title} {\bibinfo {title} {Does scrambling equal
  chaos?},\ }\href {https://doi.org/10.1103/PhysRevLett.124.140602} {\bibfield
  {journal} {\bibinfo  {journal} {Phys. Rev. Lett.}\ }\textbf {\bibinfo
  {volume} {124}},\ \bibinfo {pages} {140602} (\bibinfo {year}
  {2020})}\BibitemShut {NoStop}%
\bibitem [{\citenamefont {Maldacena}\ and\ \citenamefont
  {Stanford}(2016{\natexlab{a}})}]{PhysRevD.94.106002}%
  \BibitemOpen
  \bibfield  {author} {\bibinfo {author} {\bibfnamefont {J.}~\bibnamefont
  {Maldacena}}\ and\ \bibinfo {author} {\bibfnamefont {D.}~\bibnamefont
  {Stanford}},\ }\bibfield  {title} {\bibinfo {title} {Remarks on the
  sachdev-ye-kitaev model},\ }\href
  {https://doi.org/10.1103/PhysRevD.94.106002} {\bibfield  {journal} {\bibinfo
  {journal} {Phys. Rev. D}\ }\textbf {\bibinfo {volume} {94}},\ \bibinfo
  {pages} {106002} (\bibinfo {year} {2016}{\natexlab{a}})}\BibitemShut
  {NoStop}%
\bibitem [{\citenamefont {Iyoda}\ and\ \citenamefont
  {Sagawa}(2018)}]{Iyoda2018PRA}%
  \BibitemOpen
  \bibfield  {author} {\bibinfo {author} {\bibfnamefont {E.}~\bibnamefont
  {Iyoda}}\ and\ \bibinfo {author} {\bibfnamefont {T.}~\bibnamefont {Sagawa}},\
  }\bibfield  {title} {\bibinfo {title} {Scrambling of quantum information in
  quantum many-body systems},\ }\href
  {https://doi.org/10.1103/PhysRevA.97.042330} {\bibfield  {journal} {\bibinfo
  {journal} {Phys. Rev. A}\ }\textbf {\bibinfo {volume} {97}},\ \bibinfo
  {pages} {042330} (\bibinfo {year} {2018})}\BibitemShut {NoStop}%
\bibitem [{\citenamefont {Sahu}\ \emph {et~al.}(2019)\citenamefont {Sahu},
  \citenamefont {Xu},\ and\ \citenamefont {Swingle}}]{Sahu2019PRL}%
  \BibitemOpen
  \bibfield  {author} {\bibinfo {author} {\bibfnamefont {S.}~\bibnamefont
  {Sahu}}, \bibinfo {author} {\bibfnamefont {S.}~\bibnamefont {Xu}},\ and\
  \bibinfo {author} {\bibfnamefont {B.}~\bibnamefont {Swingle}},\ }\bibfield
  {title} {\bibinfo {title} {Scrambling dynamics across a
  thermalization-localization quantum phase transition},\ }\href
  {https://doi.org/10.1103/PhysRevLett.123.165902} {\bibfield  {journal}
  {\bibinfo  {journal} {Phys. Rev. Lett.}\ }\textbf {\bibinfo {volume} {123}},\
  \bibinfo {pages} {165902} (\bibinfo {year} {2019})}\BibitemShut {NoStop}%
\bibitem [{\citenamefont {Sahu}\ and\ \citenamefont
  {Swingle}(2020)}]{Sahu2020PRB}%
  \BibitemOpen
  \bibfield  {author} {\bibinfo {author} {\bibfnamefont {S.}~\bibnamefont
  {Sahu}}\ and\ \bibinfo {author} {\bibfnamefont {B.}~\bibnamefont {Swingle}},\
  }\bibfield  {title} {\bibinfo {title} {Information scrambling at finite
  temperature in local quantum systems},\ }\href
  {https://doi.org/10.1103/PhysRevB.102.184303} {\bibfield  {journal} {\bibinfo
   {journal} {Phys. Rev. B}\ }\textbf {\bibinfo {volume} {102}},\ \bibinfo
  {pages} {184303} (\bibinfo {year} {2020})}\BibitemShut {NoStop}%
\bibitem [{\citenamefont {Sachdev}\ and\ \citenamefont
  {Ye}(1993)}]{Sachdev1993PRL}%
  \BibitemOpen
  \bibfield  {author} {\bibinfo {author} {\bibfnamefont {S.}~\bibnamefont
  {Sachdev}}\ and\ \bibinfo {author} {\bibfnamefont {J.}~\bibnamefont {Ye}},\
  }\bibfield  {title} {\bibinfo {title} {Gapless spin-fluid ground state in a
  random quantum heisenberg magnet},\ }\href
  {https://doi.org/10.1103/PhysRevLett.70.3339} {\bibfield  {journal} {\bibinfo
   {journal} {Phys. Rev. Lett.}\ }\textbf {\bibinfo {volume} {70}},\ \bibinfo
  {pages} {3339} (\bibinfo {year} {1993})}\BibitemShut {NoStop}%
\bibitem [{\citenamefont {Maldacena}\ and\ \citenamefont
  {Stanford}(2016{\natexlab{b}})}]{Maldacena2016PRD}%
  \BibitemOpen
  \bibfield  {author} {\bibinfo {author} {\bibfnamefont {J.}~\bibnamefont
  {Maldacena}}\ and\ \bibinfo {author} {\bibfnamefont {D.}~\bibnamefont
  {Stanford}},\ }\bibfield  {title} {\bibinfo {title} {Remarks on the
  sachdev-ye-kitaev model},\ }\href
  {https://doi.org/10.1103/PhysRevD.94.106002} {\bibfield  {journal} {\bibinfo
  {journal} {Phys. Rev. D}\ }\textbf {\bibinfo {volume} {94}},\ \bibinfo
  {pages} {106002} (\bibinfo {year} {2016}{\natexlab{b}})}\BibitemShut
  {NoStop}%
\bibitem [{\citenamefont {Rosenhaus}(2019)}]{Rosenhaus2019JPA}%
  \BibitemOpen
  \bibfield  {author} {\bibinfo {author} {\bibfnamefont {V.}~\bibnamefont
  {Rosenhaus}},\ }\bibfield  {title} {\bibinfo {title} {An introduction to the
  syk model},\ }\href {https://doi.org/10.1088/1751-8121/ab2ce1} {\bibfield
  {journal} {\bibinfo  {journal} {J. Phys. A: Math. Theoret.}\ }\textbf
  {\bibinfo {volume} {52}},\ \bibinfo {pages} {323001} (\bibinfo {year}
  {2019})}\BibitemShut {NoStop}%
\bibitem [{\citenamefont {Plugge}\ \emph {et~al.}(2020)\citenamefont {Plugge},
  \citenamefont {Lantagne-Hurtubise},\ and\ \citenamefont
  {Franz}}]{Plugge2020PRL}%
  \BibitemOpen
  \bibfield  {author} {\bibinfo {author} {\bibfnamefont {S.}~\bibnamefont
  {Plugge}}, \bibinfo {author} {\bibfnamefont {E.}~\bibnamefont
  {Lantagne-Hurtubise}},\ and\ \bibinfo {author} {\bibfnamefont
  {M.}~\bibnamefont {Franz}},\ }\bibfield  {title} {\bibinfo {title} {Revival
  dynamics in a traversable wormhole},\ }\href
  {https://doi.org/10.1103/PhysRevLett.124.221601} {\bibfield  {journal}
  {\bibinfo  {journal} {Phys. Rev. Lett.}\ }\textbf {\bibinfo {volume} {124}},\
  \bibinfo {pages} {221601} (\bibinfo {year} {2020})}\BibitemShut {NoStop}%
\bibitem [{\citenamefont {Kolovsky}(2016)}]{Kolovsky2016INMP}%
  \BibitemOpen
  \bibfield  {author} {\bibinfo {author} {\bibfnamefont {A.~R.}\ \bibnamefont
  {Kolovsky}},\ }\bibfield  {title} {\bibinfo {title} {Bose–hubbard
  hamiltonian: Quantum chaos approach},\ }\href
  {https://doi.org/10.1142/S0217979216300097} {\bibfield  {journal} {\bibinfo
  {journal} {Int. J. Mod. Phys. B}\ }\textbf {\bibinfo {volume} {30}},\
  \bibinfo {pages} {1630009} (\bibinfo {year} {2016})}\BibitemShut {NoStop}%
\bibitem [{\citenamefont {Shen}\ \emph {et~al.}(2017)\citenamefont {Shen},
  \citenamefont {Zhang}, \citenamefont {Fan},\ and\ \citenamefont
  {Zhai}}]{Shen2017PRB}%
  \BibitemOpen
  \bibfield  {author} {\bibinfo {author} {\bibfnamefont {H.}~\bibnamefont
  {Shen}}, \bibinfo {author} {\bibfnamefont {P.}~\bibnamefont {Zhang}},
  \bibinfo {author} {\bibfnamefont {R.}~\bibnamefont {Fan}},\ and\ \bibinfo
  {author} {\bibfnamefont {H.}~\bibnamefont {Zhai}},\ }\bibfield  {title}
  {\bibinfo {title} {Out-of-time-order correlation at a quantum phase
  transition},\ }\href {https://doi.org/10.1103/PhysRevB.96.054503} {\bibfield
  {journal} {\bibinfo  {journal} {Phys. Rev. B}\ }\textbf {\bibinfo {volume}
  {96}},\ \bibinfo {pages} {054503} (\bibinfo {year} {2017})}\BibitemShut
  {NoStop}%
\bibitem [{\citenamefont {Freericks}\ and\ \citenamefont
  {Monien}(1994)}]{J.K.Freericks_1994}%
  \BibitemOpen
  \bibfield  {author} {\bibinfo {author} {\bibfnamefont {J.~K.}\ \bibnamefont
  {Freericks}}\ and\ \bibinfo {author} {\bibfnamefont {H.}~\bibnamefont
  {Monien}},\ }\bibfield  {title} {\bibinfo {title} {Phase diagram of the
  bose-hubbard model},\ }\href {https://doi.org/10.1209/0295-5075/26/7/012}
  {\bibfield  {journal} {\bibinfo  {journal} {EPL (Europhysics Letters)}\
  }\textbf {\bibinfo {volume} {26}},\ \bibinfo {pages} {545} (\bibinfo {year}
  {1994})}\BibitemShut {NoStop}%
\bibitem [{\citenamefont {Gardas}\ \emph {et~al.}(2017)\citenamefont {Gardas},
  \citenamefont {Dziarmaga},\ and\ \citenamefont {Zurek}}]{PhysRevB.95.104306}%
  \BibitemOpen
  \bibfield  {author} {\bibinfo {author} {\bibfnamefont {B.}~\bibnamefont
  {Gardas}}, \bibinfo {author} {\bibfnamefont {J.}~\bibnamefont {Dziarmaga}},\
  and\ \bibinfo {author} {\bibfnamefont {W.~H.}\ \bibnamefont {Zurek}},\
  }\bibfield  {title} {\bibinfo {title} {Dynamics of the quantum phase
  transition in the one-dimensional bose-hubbard model: Excitations and
  correlations induced by a quench},\ }\href
  {https://doi.org/10.1103/PhysRevB.95.104306} {\bibfield  {journal} {\bibinfo
  {journal} {Phys. Rev. B}\ }\textbf {\bibinfo {volume} {95}},\ \bibinfo
  {pages} {104306} (\bibinfo {year} {2017})}\BibitemShut {NoStop}%
\bibitem [{\citenamefont {Dziarmaga}\ and\ \citenamefont
  {Mazur}(2023)}]{PhysRevB.107.144510}%
  \BibitemOpen
  \bibfield  {author} {\bibinfo {author} {\bibfnamefont {J.}~\bibnamefont
  {Dziarmaga}}\ and\ \bibinfo {author} {\bibfnamefont {J.~M.}\ \bibnamefont
  {Mazur}},\ }\bibfield  {title} {\bibinfo {title} {Tensor network simulation
  of the quantum kibble-zurek quench from the mott to the superfluid phase in
  the two-dimensional bose-hubbard model},\ }\href
  {https://doi.org/10.1103/PhysRevB.107.144510} {\bibfield  {journal} {\bibinfo
   {journal} {Phys. Rev. B}\ }\textbf {\bibinfo {volume} {107}},\ \bibinfo
  {pages} {144510} (\bibinfo {year} {2023})}\BibitemShut {NoStop}%
\bibitem [{\citenamefont {Wang}\ and\ \citenamefont {Jiang}(2016)}]{Wang_2016}%
  \BibitemOpen
  \bibfield  {author} {\bibinfo {author} {\bibfnamefont {B.}~\bibnamefont
  {Wang}}\ and\ \bibinfo {author} {\bibfnamefont {Y.}~\bibnamefont {Jiang}},\
  }\bibfield  {title} {\bibinfo {title} {Bogoliubov approach to superfluid-bose
  glass phase transition of a disordered bose-hubbard model in weakly
  interacting regime},\ }\bibfield  {journal} {\bibinfo  {journal} {The
  European Physical Journal D}\ }\textbf {\bibinfo {volume} {70}},\ \href
  {https://doi.org/10.1140/epjd/e2016-70459-y} {10.1140/epjd/e2016-70459-y}
  (\bibinfo {year} {2016})\BibitemShut {NoStop}%
\bibitem [{\citenamefont {Kollath}\ \emph {et~al.}(2007)\citenamefont
  {Kollath}, \citenamefont {L\"auchli},\ and\ \citenamefont
  {Altman}}]{Kollath2007PRL}%
  \BibitemOpen
  \bibfield  {author} {\bibinfo {author} {\bibfnamefont {C.}~\bibnamefont
  {Kollath}}, \bibinfo {author} {\bibfnamefont {A.~M.}\ \bibnamefont
  {L\"auchli}},\ and\ \bibinfo {author} {\bibfnamefont {E.}~\bibnamefont
  {Altman}},\ }\bibfield  {title} {\bibinfo {title} {Quench dynamics and
  nonequilibrium phase diagram of the bose-hubbard model},\ }\href
  {https://doi.org/10.1103/PhysRevLett.98.180601} {\bibfield  {journal}
  {\bibinfo  {journal} {Phys. Rev. Lett.}\ }\textbf {\bibinfo {volume} {98}},\
  \bibinfo {pages} {180601} (\bibinfo {year} {2007})}\BibitemShut {NoStop}%
\bibitem [{\citenamefont {Nakerst}\ and\ \citenamefont
  {Haque}(2023)}]{Goran2023PRE}%
  \BibitemOpen
  \bibfield  {author} {\bibinfo {author} {\bibfnamefont {G.}~\bibnamefont
  {Nakerst}}\ and\ \bibinfo {author} {\bibfnamefont {M.}~\bibnamefont
  {Haque}},\ }\bibfield  {title} {\bibinfo {title} {Chaos in the three-site
  bose-hubbard model: Classical versus quantum},\ }\href
  {https://doi.org/10.1103/PhysRevE.107.024210} {\bibfield  {journal} {\bibinfo
   {journal} {Phys. Rev. E}\ }\textbf {\bibinfo {volume} {107}},\ \bibinfo
  {pages} {024210} (\bibinfo {year} {2023})}\BibitemShut {NoStop}%
\bibitem [{\citenamefont {T.O.~Wehling}\ and\ \citenamefont
  {Balatsky}(2014)}]{Wehling2014AP}%
  \BibitemOpen
  \bibfield  {author} {\bibinfo {author} {\bibfnamefont {A.~B.-S.}\
  \bibnamefont {T.O.~Wehling}}\ and\ \bibinfo {author} {\bibfnamefont
  {A.}~\bibnamefont {Balatsky}},\ }\bibfield  {title} {\bibinfo {title} {Dirac
  materials},\ }\href {https://doi.org/10.1080/00018732.2014.927109} {\bibfield
   {journal} {\bibinfo  {journal} {Ad. Phys.}\ }\textbf {\bibinfo {volume}
  {63}},\ \bibinfo {pages} {1} (\bibinfo {year} {2014})}\BibitemShut {NoStop}%
\bibitem [{\citenamefont {Gonz\'alez~Alonso}\ \emph {et~al.}(2019)\citenamefont
  {Gonz\'alez~Alonso}, \citenamefont {Yunger~Halpern},\ and\ \citenamefont
  {Dressel}}]{PhysRevLett.122.040404}%
  \BibitemOpen
  \bibfield  {author} {\bibinfo {author} {\bibfnamefont {J.~R.}\ \bibnamefont
  {Gonz\'alez~Alonso}}, \bibinfo {author} {\bibfnamefont {N.}~\bibnamefont
  {Yunger~Halpern}},\ and\ \bibinfo {author} {\bibfnamefont {J.}~\bibnamefont
  {Dressel}},\ }\bibfield  {title} {\bibinfo {title}
  {Out-of-time-ordered-correlator quasiprobabilities robustly witness
  scrambling},\ }\href {https://doi.org/10.1103/PhysRevLett.122.040404}
  {\bibfield  {journal} {\bibinfo  {journal} {Phys. Rev. Lett.}\ }\textbf
  {\bibinfo {volume} {122}},\ \bibinfo {pages} {040404} (\bibinfo {year}
  {2019})}\BibitemShut {NoStop}%
\bibitem [{\citenamefont {Yan}\ and\ \citenamefont {Zurek}(2022)}]{Yan2022NJP}%
  \BibitemOpen
  \bibfield  {author} {\bibinfo {author} {\bibfnamefont {B.}~\bibnamefont
  {Yan}}\ and\ \bibinfo {author} {\bibfnamefont {W.~H.}\ \bibnamefont
  {Zurek}},\ }\bibfield  {title} {\bibinfo {title} {Decoherence factor as a
  convolution: an interplay between a gaussian and an exponential coherence
  loss},\ }\href {https://doi.org/10.1088/1367-2630/ac9fe8} {\bibfield
  {journal} {\bibinfo  {journal} {New J. Phys.}\ }\textbf {\bibinfo {volume}
  {24}},\ \bibinfo {pages} {113029} (\bibinfo {year} {2022})}\BibitemShut
  {NoStop}%
\bibitem [{\citenamefont {Arovas}\ \emph {et~al.}(2022)\citenamefont {Arovas},
  \citenamefont {Berg}, \citenamefont {Kivelson},\ and\ \citenamefont
  {Raghu}}]{Arovas2022ARCMP}%
  \BibitemOpen
  \bibfield  {author} {\bibinfo {author} {\bibfnamefont {D.~P.}\ \bibnamefont
  {Arovas}}, \bibinfo {author} {\bibfnamefont {E.}~\bibnamefont {Berg}},
  \bibinfo {author} {\bibfnamefont {S.~A.}\ \bibnamefont {Kivelson}},\ and\
  \bibinfo {author} {\bibfnamefont {S.}~\bibnamefont {Raghu}},\ }\bibfield
  {title} {\bibinfo {title} {The hubbard model},\ }\href
  {https://doi.org/10.1146/annurev-conmatphys-031620-102024} {\bibfield
  {journal} {\bibinfo  {journal} {Ann. Rev. Cond. Matt. Phys.}\ }\textbf
  {\bibinfo {volume} {13}},\ \bibinfo {pages} {239} (\bibinfo {year}
  {2022})}\BibitemShut {NoStop}%
\bibitem [{\citenamefont {Dutta}\ \emph {et~al.}(2015)\citenamefont {Dutta},
  \citenamefont {Gajda}, \citenamefont {Hauke}, \citenamefont {Lewenstein},
  \citenamefont {Lühmann}, \citenamefont {Malomed}, \citenamefont
  {Sowiński},\ and\ \citenamefont {Zakrzewski}}]{Dutta2015RPP}%
  \BibitemOpen
  \bibfield  {author} {\bibinfo {author} {\bibfnamefont {O.}~\bibnamefont
  {Dutta}}, \bibinfo {author} {\bibfnamefont {M.}~\bibnamefont {Gajda}},
  \bibinfo {author} {\bibfnamefont {P.}~\bibnamefont {Hauke}}, \bibinfo
  {author} {\bibfnamefont {M.}~\bibnamefont {Lewenstein}}, \bibinfo {author}
  {\bibfnamefont {D.-S.}\ \bibnamefont {Lühmann}}, \bibinfo {author}
  {\bibfnamefont {B.~A.}\ \bibnamefont {Malomed}}, \bibinfo {author}
  {\bibfnamefont {T.}~\bibnamefont {Sowiński}},\ and\ \bibinfo {author}
  {\bibfnamefont {J.}~\bibnamefont {Zakrzewski}},\ }\bibfield  {title}
  {\bibinfo {title} {Non-standard hubbard models in optical lattices: a
  review},\ }\href {https://doi.org/10.1088/0034-4885/78/6/066001} {\bibfield
  {journal} {\bibinfo  {journal} {Rep. Prog. Phys.}\ }\textbf {\bibinfo
  {volume} {78}},\ \bibinfo {pages} {066001} (\bibinfo {year}
  {2015})}\BibitemShut {NoStop}%
\bibitem [{\citenamefont {Greiner}\ \emph {et~al.}(2002)\citenamefont
  {Greiner}, \citenamefont {Mandel}, \citenamefont {Esslinger}, \citenamefont
  {H{\"a}nsch},\ and\ \citenamefont {Bloch}}]{Greiner2002}%
  \BibitemOpen
  \bibfield  {author} {\bibinfo {author} {\bibfnamefont {M.}~\bibnamefont
  {Greiner}}, \bibinfo {author} {\bibfnamefont {O.}~\bibnamefont {Mandel}},
  \bibinfo {author} {\bibfnamefont {T.}~\bibnamefont {Esslinger}}, \bibinfo
  {author} {\bibfnamefont {T.~W.}\ \bibnamefont {H{\"a}nsch}},\ and\ \bibinfo
  {author} {\bibfnamefont {I.}~\bibnamefont {Bloch}},\ }\bibfield  {title}
  {\bibinfo {title} {Quantum phase transition from a superfluid to a mott
  insulator in a gas of ultracold atoms},\ }\href
  {https://doi.org/10.1038/415039a} {\bibfield  {journal} {\bibinfo  {journal}
  {Nature}\ }\textbf {\bibinfo {volume} {415}},\ \bibinfo {pages} {39}
  (\bibinfo {year} {2002})}\BibitemShut {NoStop}%
\bibitem [{\citenamefont {Dziarmaga}\ and\ \citenamefont
  {Zurek}(2014)}]{Dziarmaga2014}%
  \BibitemOpen
  \bibfield  {author} {\bibinfo {author} {\bibfnamefont {J.}~\bibnamefont
  {Dziarmaga}}\ and\ \bibinfo {author} {\bibfnamefont {W.~H.}\ \bibnamefont
  {Zurek}},\ }\bibfield  {title} {\bibinfo {title} {Quench in the 1d
  bose-hubbard model: Topological defects and excitations from the
  kosterlitz-thouless phase transition dynamics},\ }\href
  {https://doi.org/10.1038/srep05950} {\bibfield  {journal} {\bibinfo
  {journal} {Scientific Reports}\ }\textbf {\bibinfo {volume} {4}},\ \bibinfo
  {pages} {5950} (\bibinfo {year} {2014})}\BibitemShut {NoStop}%
\bibitem [{\citenamefont {Shimizu}\ \emph {et~al.}(2018)\citenamefont
  {Shimizu}, \citenamefont {Kuno}, \citenamefont {Hirano},\ and\ \citenamefont
  {Ichinose}}]{PhysRevA.97.033626}%
  \BibitemOpen
  \bibfield  {author} {\bibinfo {author} {\bibfnamefont {K.}~\bibnamefont
  {Shimizu}}, \bibinfo {author} {\bibfnamefont {Y.}~\bibnamefont {Kuno}},
  \bibinfo {author} {\bibfnamefont {T.}~\bibnamefont {Hirano}},\ and\ \bibinfo
  {author} {\bibfnamefont {I.}~\bibnamefont {Ichinose}},\ }\bibfield  {title}
  {\bibinfo {title} {Dynamics of a quantum phase transition in the bose-hubbard
  model: Kibble-zurek mechanism and beyond},\ }\href
  {https://doi.org/10.1103/PhysRevA.97.033626} {\bibfield  {journal} {\bibinfo
  {journal} {Phys. Rev. A}\ }\textbf {\bibinfo {volume} {97}},\ \bibinfo
  {pages} {033626} (\bibinfo {year} {2018})}\BibitemShut {NoStop}%
\bibitem [{\citenamefont {Weiss}\ \emph {et~al.}(2018)\citenamefont {Weiss},
  \citenamefont {Gerster}, \citenamefont {Jaschke}, \citenamefont {Silvi},\
  and\ \citenamefont {Montangero}}]{PhysRevA.98.063601}%
  \BibitemOpen
  \bibfield  {author} {\bibinfo {author} {\bibfnamefont {W.}~\bibnamefont
  {Weiss}}, \bibinfo {author} {\bibfnamefont {M.}~\bibnamefont {Gerster}},
  \bibinfo {author} {\bibfnamefont {D.}~\bibnamefont {Jaschke}}, \bibinfo
  {author} {\bibfnamefont {P.}~\bibnamefont {Silvi}},\ and\ \bibinfo {author}
  {\bibfnamefont {S.}~\bibnamefont {Montangero}},\ }\bibfield  {title}
  {\bibinfo {title} {Kibble-zurek scaling of the one-dimensional bose-hubbard
  model at finite temperatures},\ }\href
  {https://doi.org/10.1103/PhysRevA.98.063601} {\bibfield  {journal} {\bibinfo
  {journal} {Phys. Rev. A}\ }\textbf {\bibinfo {volume} {98}},\ \bibinfo
  {pages} {063601} (\bibinfo {year} {2018})}\BibitemShut {NoStop}%
\bibitem [{\citenamefont {Goldstein}\ \emph {et~al.}(2002)\citenamefont
  {Goldstein}, \citenamefont {Poole},\ and\ \citenamefont
  {Safko}}]{Goldstein2002}%
  \BibitemOpen
  \bibfield  {author} {\bibinfo {author} {\bibfnamefont {H.}~\bibnamefont
  {Goldstein}}, \bibinfo {author} {\bibfnamefont {C.}~\bibnamefont {Poole}},\
  and\ \bibinfo {author} {\bibfnamefont {J.}~\bibnamefont {Safko}},\
  }\href@noop {} {\emph {\bibinfo {title} {Classical mechanics}}}\ (\bibinfo
  {publisher} {American Association of Physics Teachers},\ \bibinfo {year}
  {2002})\BibitemShut {NoStop}%
\bibitem [{\citenamefont {Shenker}\ and\ \citenamefont
  {Stanford}(2014)}]{shenker2014black}%
  \BibitemOpen
  \bibfield  {author} {\bibinfo {author} {\bibfnamefont {S.~H.}\ \bibnamefont
  {Shenker}}\ and\ \bibinfo {author} {\bibfnamefont {D.}~\bibnamefont
  {Stanford}},\ }\bibfield  {title} {\bibinfo {title} {Black holes and the
  butterfly effect},\ }\href@noop {} {\bibfield  {journal} {\bibinfo  {journal}
  {JHEP}\ }\textbf {\bibinfo {volume} {2014}}\bibinfo  {number} { (3)},\
  \bibinfo {pages} {1}}\BibitemShut {NoStop}%
\bibitem [{\citenamefont {Gong}\ \emph {et~al.}(2023)\citenamefont {Gong},
  \citenamefont {Guaita},\ and\ \citenamefont {Cirac}}]{Gong2023PRL}%
  \BibitemOpen
\bibfield  {number} {  }\bibfield  {author} {\bibinfo {author} {\bibfnamefont
  {Z.}~\bibnamefont {Gong}}, \bibinfo {author} {\bibfnamefont {T.}~\bibnamefont
  {Guaita}},\ and\ \bibinfo {author} {\bibfnamefont {J.~I.}\ \bibnamefont
  {Cirac}},\ }\bibfield  {title} {\bibinfo {title} {Long-range free fermions:
  Lieb-robinson bound, clustering properties, and topological phases},\ }\href
  {https://doi.org/10.1103/PhysRevLett.130.070401} {\bibfield  {journal}
  {\bibinfo  {journal} {Phys. Rev. Lett.}\ }\textbf {\bibinfo {volume} {130}},\
  \bibinfo {pages} {070401} (\bibinfo {year} {2023})}\BibitemShut {NoStop}%
\bibitem [{\citenamefont {Nachtergaele}\ \emph {et~al.}(2009)\citenamefont
  {Nachtergaele}, \citenamefont {Raz}, \citenamefont {Schlein},\ and\
  \citenamefont {Sims}}]{Nachtergaele2009}%
  \BibitemOpen
  \bibfield  {author} {\bibinfo {author} {\bibfnamefont {B.}~\bibnamefont
  {Nachtergaele}}, \bibinfo {author} {\bibfnamefont {H.}~\bibnamefont {Raz}},
  \bibinfo {author} {\bibfnamefont {B.}~\bibnamefont {Schlein}},\ and\ \bibinfo
  {author} {\bibfnamefont {R.}~\bibnamefont {Sims}},\ }\bibfield  {title}
  {\bibinfo {title} {Lieb-robinson bounds for harmonic and anharmonic lattice
  systems},\ }\href {https://doi.org/10.1007/s00220-008-0630-2} {\bibfield
  {journal} {\bibinfo  {journal} {Comm. Math. Phys.}\ }\textbf {\bibinfo
  {volume} {286}},\ \bibinfo {pages} {1073} (\bibinfo {year}
  {2009})}\BibitemShut {NoStop}%
\bibitem [{\citenamefont {Hastings}(2010)}]{hastings2010locality}%
  \BibitemOpen
  \bibfield  {author} {\bibinfo {author} {\bibfnamefont {M.~B.}\ \bibnamefont
  {Hastings}},\ }\href@noop {} {\bibinfo {title} {Locality in quantum systems}}
  (\bibinfo {year} {2010}),\ \Eprint {https://arxiv.org/abs/1008.5137}
  {arXiv:1008.5137 [math-ph]} \BibitemShut {NoStop}%
\bibitem [{\citenamefont {Roberts}\ and\ \citenamefont
  {Swingle}(2016{\natexlab{b}})}]{roberts2016lieb}%
  \BibitemOpen
  \bibfield  {author} {\bibinfo {author} {\bibfnamefont {D.~A.}\ \bibnamefont
  {Roberts}}\ and\ \bibinfo {author} {\bibfnamefont {B.}~\bibnamefont
  {Swingle}},\ }\bibfield  {title} {\bibinfo {title} {Lieb-robinson bound and
  the butterfly effect in quantum field theories},\ }\href@noop {} {\bibfield
  {journal} {\bibinfo  {journal} {Physical review letters}\ }\textbf {\bibinfo
  {volume} {117}},\ \bibinfo {pages} {091602} (\bibinfo {year}
  {2016}{\natexlab{b}})}\BibitemShut {NoStop}%
\bibitem [{\citenamefont {Ford}(2015)}]{FORD2015491}%
  \BibitemOpen
  \bibfield  {author} {\bibinfo {author} {\bibfnamefont {W.}~\bibnamefont
  {Ford}},\ }\bibfield  {title} {\bibinfo {title} {Chapter 21 - krylov subspace
  methods},\ }in\ \href
  {https://doi.org/https://doi.org/10.1016/B978-0-12-394435-1.00021-1} {\emph
  {\bibinfo {booktitle} {Numerical Linear Algebra with Applications}}},\
  \bibinfo {editor} {edited by\ \bibinfo {editor} {\bibfnamefont
  {W.}~\bibnamefont {Ford}}}\ (\bibinfo  {publisher} {Academic Press},\
  \bibinfo {address} {Boston},\ \bibinfo {year} {2015})\ pp.\ \bibinfo {pages}
  {491--532}\BibitemShut {NoStop}%
\bibitem [{\citenamefont {Yan}\ \emph {et~al.}(2020)\citenamefont {Yan},
  \citenamefont {Cincio},\ and\ \citenamefont {Zurek}}]{Yan2020PRL}%
  \BibitemOpen
  \bibfield  {author} {\bibinfo {author} {\bibfnamefont {B.}~\bibnamefont
  {Yan}}, \bibinfo {author} {\bibfnamefont {L.}~\bibnamefont {Cincio}},\ and\
  \bibinfo {author} {\bibfnamefont {W.~H.}\ \bibnamefont {Zurek}},\ }\bibfield
  {title} {\bibinfo {title} {Information scrambling and loschmidt echo},\
  }\href {https://doi.org/10.1103/PhysRevLett.124.160603} {\bibfield  {journal}
  {\bibinfo  {journal} {Phys. Rev. Lett.}\ }\textbf {\bibinfo {volume} {124}},\
  \bibinfo {pages} {160603} (\bibinfo {year} {2020})}\BibitemShut {NoStop}%
\bibitem [{\citenamefont {Cucchietti}(2004)}]{cucchietti2004loschmidt}%
  \BibitemOpen
  \bibfield  {author} {\bibinfo {author} {\bibfnamefont {F.~M.}\ \bibnamefont
  {Cucchietti}},\ }\href@noop {} {\bibinfo {title} {The loschmidt echo in
  classically chaotic systems: Quantum chaos, irreversibility and decoherence}}
  (\bibinfo {year} {2004}),\ \Eprint {https://arxiv.org/abs/quant-ph/0410121}
  {arXiv:quant-ph/0410121 [quant-ph]} \BibitemShut {NoStop}%
\bibitem [{\citenamefont {Gorin}\ \emph {et~al.}(2006)\citenamefont {Gorin},
  \citenamefont {Prosen}, \citenamefont {Seligman},\ and\ \citenamefont
  {Žnidarič}}]{GORIN200633}%
  \BibitemOpen
  \bibfield  {author} {\bibinfo {author} {\bibfnamefont {T.}~\bibnamefont
  {Gorin}}, \bibinfo {author} {\bibfnamefont {T.}~\bibnamefont {Prosen}},
  \bibinfo {author} {\bibfnamefont {T.~H.}\ \bibnamefont {Seligman}},\ and\
  \bibinfo {author} {\bibfnamefont {M.}~\bibnamefont {Žnidarič}},\ }\bibfield
   {title} {\bibinfo {title} {Dynamics of loschmidt echoes and fidelity
  decay},\ }\href
  {https://doi.org/https://doi.org/10.1016/j.physrep.2006.09.003} {\bibfield
  {journal} {\bibinfo  {journal} {Phys. Rep.}\ }\textbf {\bibinfo {volume}
  {435}},\ \bibinfo {pages} {33} (\bibinfo {year} {2006})}\BibitemShut
  {NoStop}%
\bibitem [{\citenamefont {Ares}\ and\ \citenamefont
  {Wisniacki}(2009)}]{PhysRevE.80.046216}%
  \BibitemOpen
  \bibfield  {author} {\bibinfo {author} {\bibfnamefont {N.}~\bibnamefont
  {Ares}}\ and\ \bibinfo {author} {\bibfnamefont {D.~A.}\ \bibnamefont
  {Wisniacki}},\ }\bibfield  {title} {\bibinfo {title} {Loschmidt echo and the
  local density of states},\ }\href
  {https://doi.org/10.1103/PhysRevE.80.046216} {\bibfield  {journal} {\bibinfo
  {journal} {Phys. Rev. E}\ }\textbf {\bibinfo {volume} {80}},\ \bibinfo
  {pages} {046216} (\bibinfo {year} {2009})}\BibitemShut {NoStop}%
\bibitem [{\citenamefont {Touil}\ and\ \citenamefont
  {Deffner}(2020)}]{Touil2020QST}%
  \BibitemOpen
  \bibfield  {author} {\bibinfo {author} {\bibfnamefont {A.}~\bibnamefont
  {Touil}}\ and\ \bibinfo {author} {\bibfnamefont {S.}~\bibnamefont
  {Deffner}},\ }\bibfield  {title} {\bibinfo {title} {Quantum scrambling and
  the growth of mutual information},\ }\href
  {https://doi.org/10.1088/2058-9565/ab8ebb} {\bibfield  {journal} {\bibinfo
  {journal} {Quantum Science and Technology}\ }\textbf {\bibinfo {volume}
  {5}},\ \bibinfo {pages} {035005} (\bibinfo {year} {2020})}\BibitemShut
  {NoStop}%
\bibitem [{\citenamefont {Seshadri}\ \emph {et~al.}(2018)\citenamefont
  {Seshadri}, \citenamefont {Madhok},\ and\ \citenamefont
  {Lakshminarayan}}]{PhysRevE.98.052205}%
  \BibitemOpen
  \bibfield  {author} {\bibinfo {author} {\bibfnamefont {A.}~\bibnamefont
  {Seshadri}}, \bibinfo {author} {\bibfnamefont {V.}~\bibnamefont {Madhok}},\
  and\ \bibinfo {author} {\bibfnamefont {A.}~\bibnamefont {Lakshminarayan}},\
  }\bibfield  {title} {\bibinfo {title} {Tripartite mutual information,
  entanglement, and scrambling in permutation symmetric systems with an
  application to quantum chaos},\ }\href
  {https://doi.org/10.1103/PhysRevE.98.052205} {\bibfield  {journal} {\bibinfo
  {journal} {Phys. Rev. E}\ }\textbf {\bibinfo {volume} {98}},\ \bibinfo
  {pages} {052205} (\bibinfo {year} {2018})}\BibitemShut {NoStop}%
\bibitem [{\citenamefont {Hosur}\ \emph {et~al.}(2016)\citenamefont {Hosur},
  \citenamefont {Qi}, \citenamefont {Roberts},\ and\ \citenamefont
  {Yoshida}}]{Hosur2016}%
  \BibitemOpen
  \bibfield  {author} {\bibinfo {author} {\bibfnamefont {P.}~\bibnamefont
  {Hosur}}, \bibinfo {author} {\bibfnamefont {X.-L.}\ \bibnamefont {Qi}},
  \bibinfo {author} {\bibfnamefont {D.~A.}\ \bibnamefont {Roberts}},\ and\
  \bibinfo {author} {\bibfnamefont {B.}~\bibnamefont {Yoshida}},\ }\bibfield
  {title} {\bibinfo {title} {Chaos in quantum channels},\ }\href
  {https://doi.org/10.1007/JHEP02(2016)004} {\bibfield  {journal} {\bibinfo
  {journal} {JHEP}\ }\textbf {\bibinfo {volume} {2016}}\bibinfo  {number} {
  (2)},\ \bibinfo {pages} {4}}\BibitemShut {NoStop}%
\bibitem [{\citenamefont {Han}\ \emph {et~al.}(2022)\citenamefont {Han},
  \citenamefont {Zou}, \citenamefont {Li},\ and\ \citenamefont
  {Shao}}]{e24111532}%
  \BibitemOpen
\bibfield  {number} {  }\bibfield  {author} {\bibinfo {author} {\bibfnamefont
  {L.-P.}\ \bibnamefont {Han}}, \bibinfo {author} {\bibfnamefont
  {J.}~\bibnamefont {Zou}}, \bibinfo {author} {\bibfnamefont {H.}~\bibnamefont
  {Li}},\ and\ \bibinfo {author} {\bibfnamefont {B.}~\bibnamefont {Shao}},\
  }\bibfield  {title} {\bibinfo {title} {Quantum information scrambling in
  non-markovian open quantum system},\ }\bibfield  {journal} {\bibinfo
  {journal} {Entropy}\ }\textbf {\bibinfo {volume} {24}},\ \href
  {https://doi.org/10.3390/e24111532} {10.3390/e24111532} (\bibinfo {year}
  {2022})\BibitemShut {NoStop}%
\end{thebibliography}%

\end{document}